\pdfoutput=1
\documentclass[fleqn, usenatbib]{mnras}

\usepackage{newtxtext,newtxmath}
\usepackage[T1]{fontenc}

\DeclareRobustCommand{\VAN}[3]{#2}
\let\VANthebibliography\thebibliography
\def\thebibliography{\DeclareRobustCommand{\VAN}[3]{##3}\VANthebibliography}

\usepackage{graphicx}
\usepackage{amsmath}
\usepackage{subcaption}
\usepackage{booktabs}
\usepackage{xspace}
\usepackage{xcolor}
\usepackage{hyperref}
\usepackage{orcidlink}

\makeatletter
\newenvironment{placedfigure}{%
    \par\vbox\bgroup
    \hsize=\columnwidth
    \captionsetup{type=figure}%
    \let\@makecaption\SFB@makefigurecaption
    \fstyle@figure
}{\par\egroup\medskip}
\makeatother

\newcommand{\dsct}{$\delta$\,Sct\xspace}
\newcommand{\mesa}{\textsc{mesa}\xspace}
\newcommand{\gyre}{\textsc{gyre}\xspace}
\newcommand{\storm}{\textsc{storm}\xspace}
\newcommand{\ultranest}{\textsc{UltraNest}\xspace}

\title[NN emulator for \dsct asteroseismology]{An asteroseismic age estimate for the Pleiades using a neural-network emulator of rotating $\delta$\,Scuti pulsation models}

\author[Gautam et al.]{
Anuj Gautam\,\orcidlink{0000-0003-0619-0207}$^{1}$\thanks{E-mail: anuj.gautam@unisq.edu.au (AG)}
and Simon J. Murphy\,\orcidlink{0000-0002-5648-3107}$^{1}$\thanks{E-mail: simon.murphy@unisq.edu.au (SJM)}
\\
$^{1}$ Centre for Astrophysics, University of Southern Queensland, Toowoomba, QLD 4350, Australia\\
}

\date{Accepted XXX. Received YYY; in original form ZZZ}
\pubyear{YYYY}

\begin{document}
\label{firstpage}
\pagerange{\pageref{firstpage}--\PreviousTotalPages}
\maketitle

\begin{abstract}
    Asteroseismology of \dsct stars has strong potential for determining the ages of young stars through individual mode fitting. We present a neural-network emulator trained on approximately 46 million rotating pulsation models. From stellar mass, metallicity, age, and equatorial rotation velocity, it predicts 12 classical observables, two asteroseismic scalars, and the p\:mode, f\:mode and low-order g\:mode frequencies for degrees $\ell=0$--3, including the individual $m$ components of rotationally split multiplets. We couple the emulator to nested sampling through a multivariate likelihood that propagates correlated emulator residuals, making full Bayesian posterior exploration practical. Applied to five \dsct stars in the Pleiades observed by \textit{TESS}, the framework gives mutually consistent individual asteroseismic ages of 107--117\,Myr. Combining their age posteriors gives a cluster age of $113.3\pm4.5$\,Myr (statistical), consistent with the $112 \pm 5$\,Myr lithium-depletion-boundary age and the 110--160\,Myr range from rotating-isochrone fitting. Together with spectroscopic $v\sin i$, the inferred equatorial velocities also provide post-fit inclination estimates. Separate perturbed-grid tests using 100 matched synthetic pairs per input-physics variation give median absolute age shifts of $2.1$, $2.3$, and $5.2$\,Myr for $|\Delta f_{\rm ov}|=0.005$, $|\Delta\alpha_{\rm MLT}|=0.1$, and $|\Delta Y|=0.02$, respectively; the corresponding 90th percentile shifts are $6.5$, $7.6$, and $16.2$\,Myr. These tests describe conditional population sensitivities, not a systematic uncertainty on the Pleiades age. Together, the cluster-age agreement and the scale of these input-physics sensitivities support \dsct asteroseismology as a complementary age estimation method for young stars.
\end{abstract}


\section{Introduction}
\label{sec:intro}

The pulsations of \dsct stars offer a practical route to their ages. They are primarily low-overtone p\:modes driven principally by the $\kappa$-mechanism in the He\,\textsc{ii} ionisation zone \citep{pamyatnykh_pulsational_1999, aerts_asteroseismology_2010}, with envelope mixing and turbulent pressure also contributing to the observed extent of the instability strip \citep{antoci_role_2014, antoci_first_2019}. Their frequencies sample the interior sound-speed and density structure, and therefore the evolutionary phase \citep{aerts_age_2015, kurtz_asteroseismology_2022}. The pulsation spectra of \dsct stars are difficult to interpret because only some modes are excited, an unpredictable subset of which reaches detectable amplitude \citep{breger_problems_2002, smolec_mode_2013}, and rotation shifts the frequencies of the ones that do. \citet{bedding_very_2020} showed that some young \dsct stars nonetheless exhibit ordered high-frequency pulsations in which radial and dipole ridges can be identified in an \'echelle diagram. With modes identified, frequencies can be fitted to grids of pulsation models \citep{murphy_five_2022, murphy_grid_2023}. For the best-constrained stars the resulting age precisions are better than 10\,per\,cent: \citet{murphy_precise_2021} obtained $10.75 \pm 0.77$\,Myr for HD\,139614, and such ages have since been incorporated into the star formation histories of young associations \citep{kerr_spyglass_2022}.

The mass range occupied by \dsct stars, roughly 1.5--2.3\,M$_\odot$ \citep{murphyPopulationSynthesisDelta2026b}, is where other age methods provide relatively weak constraints. Ages there rest almost entirely on isochrone placement \citep{david_ages_2015}. Isochrones lose sensitivity for stars still close to the zero-age main-sequence \citep[$\lesssim$ a third of the main-sequence lifetime;][]{soderblom_ages_2010}. Population synthesis of intermediate-mass stars finds ages underestimated by a factor of two or more near the zero-age main sequence, and overestimated by $31$\,per\,cent once a star is past a tenth of its main-sequence lifetime \citep{murphyQuantifyingIsochronebasedAge2026}. Colour--magnitude isochrone fitting is most sensitive near the main-sequence turn-off, where rotation inflates the age spread inferred from non-rotating models \citep{lipatov_rotational_2022}. Isochrone placement also leaves age degenerate with composition, because isochrones of different ages and metallicities can occupy the same point in the Hertzsprung--Russell diagram. Other methods such as the lithium-depletion boundary and kinematic traceback instead need low-mass members, so they work only for clusters and associations \citep{galindo-guil_lithium_2022, miret-roig_star_2022, zerjal_chronostar_2023}. Assigning an intermediate-mass star the age of its cluster's low-mass members presumes coevality across mass rather than establishing it, and a young field star has no cluster age to inherit at all. The gap in reliable ages for intermediate-mass stars matters because protoplanetary discs disperse faster above about 2\,M$_\odot$ \citep{ribas_protoplanetary_2015}, and giant-planet occurrence peaks near the same mass before falling away \citep{reffert_precise_2015, wolthoff_precise_2022}. A shorter disc lifetime has been invoked to explain both trends, but testing that explanation requires ages in this mass range. 

Cluster asteroseismology has focused mostly on red giants \citep{basu_sounding_2011}, while main-sequence A stars have received far less attention. That is changing now that \textit{TESS} covers most of the sky. Individual \dsct pulsators have been characterised in the Pleiades \citep{murphy_five_2022, bedding_pleiades_2023}, $\alpha$\,Per \citep{pamos-ortega_seismic_2022}, Praesepe \citep{pamos-ortega_dating_2023}, the Cep--Her Complex \citep{murphy_scuti_2024} and NGC\,3532 \citep{berry_discovery_2025}. In addition to ages, near-core rotation rates have also been inferred from gravity-mode pulsators in NGC\,3532 and NGC\,2516 \citep{li_gravity-mode_2026}.

The Pleiades is a natural benchmark for age estimation methods because it is already well studied, allowing a new determination to be compared with existing ones. The lithium-depletion boundary gives $112 \pm 5$\,Myr \citep{dahm_2015}, revised downward from the $125 \pm 8$\,Myr of \citet{stauffer_pleiades_1998} as the underlying models were updated. Rotating \mesa isochrones fitted to the colour--magnitude diagram span roughly 110--160\,Myr \citep{gossage_pleiades_2018}. The two methods are consistent, but the isochronal range is wide and the lithium-depletion-boundary value has shifted by more than its quoted uncertainty. An earlier asteroseismic study by \citet{fox_machado_seismological_2006} compared the frequencies of six Pleiades \dsct stars with rotating models at ages of 70, 100 and 130\,Myr. They inferred mode identifications during fitting, with their frequency-fitting and mode-visibility criteria favouring models aged 70--100\,Myr.

More recently, \citet{murphy_five_2022} identified pulsation modes in five Pleiades \dsct stars using K2 photometry, and \textit{TESS} has since detected \dsct pulsations in 36 members \citep{bedding_pleiades_2023}. The cluster therefore offers both an external age to compare against and enough characterised pulsators to select those whose \'echelle diagrams support secure mode identifications. We revisit its asteroseismic age with five such stars in Section~\ref{sec:pleiades}.

Three bottlenecks stand between a \dsct pulsation spectrum and an age estimate. The first is the cost of computing the models themselves. Each requires an evolutionary track followed by a pulsation calculation at each timestep. With rotation included, the latter takes seconds to minutes, and a complete track with stellar evolution and pulsation calculations can take dozens of CPU hours depending on the required precision. Our approach pays this cost once by constructing a precomputed grid on which an emulator can be trained. Such grids have grown by more than two orders of magnitude in the past few years, from the $2 \times 10^{5}$ models of \citet{murphy_grid_2023} to the $2.5 \times 10^{7}$ of \citet{gautam_modelling_2025}, hereafter Paper~I. 

The second bottleneck is inference from that grid. Bayesian inference requires comparing predicted and observed frequencies across parameter space, and a sampler requires $10^{4}$--$10^{6}$ evaluations per star at arbitrary parameter values rather than at grid points. Recomputing tracks during sampling is infeasible \citep{scutt_asteroseismology_2023}. One solution is to interpolate within the stored grid, but both the cost and the error of interpolation grow with the number of sampled dimensions \citep{scutt_pitchfork_2026}. The penalty is heavier for a grid of rotating models, because rotation adds a sampled dimension and rotational splitting multiplies the number of frequencies that must be predicted at each grid point. Achievable precision is then set partly by the grid spacing and not only by the observational uncertainties. Moreover, until the interpolation error is well controlled it cannot be separated from the systematic uncertainty of the adopted input physics. 

Machine learning offers a different route, and it is not new to asteroseismic inference. Random forests have been trained on global seismic quantities \citep{bellinger_machine_2016}, and neural networks have been used to classify pulsation spectra \citep{hon_deep_2018}. Neural networks have also been explored for estimating the masses, metallicities, and ages of \dsct stars using seismic scalars such as the large frequency separation $\Delta\nu$ and the asymptotic phase offset $\varepsilon$, or individual mode frequencies \citep{pandaAsteroseismologyAppliedConstrain2024}. As forward emulators, neural networks make it practical to include additional physics such as variable mixing \citep{lyttle_modelling_2021} or rotation \citep{mombarg_constraining_2021}, and to predict individual frequencies rather than summary statistics. \citet{scutt_asteroseismology_2023} applied this approach to \dsct stars by training a network on non-rotating models and sampling from it to obtain posteriors that reflect the random uncertainties, and \citet{scutt_pitchfork_2026} have since done the same for individual frequencies of solar-like oscillators. 

The evaluation cost of a neural-network emulator is nearly independent of the number of inputs. Its accuracy still depends on grid spacing: for grids of solar-type main-sequence models in mass and age, \citet{gluck_benchmarking_2026} find that neural networks are more accurate than linear interpolation but lose accuracy as the grid becomes sparser. The emulator's residuals can, however, be measured on held-out models and propagated into the likelihood. Measuring the emulator error in that way keeps it separate from the systematic uncertainty of the adopted input physics, which an uncontrolled interpolation error would otherwise be confounded with. We revisit the uncertainty associated with the adopted input physics in Paper~III of this series.

The third bottleneck is identifying modes in rotating stars, usually done by labelling peaks on the \'echelle diagrams of the \textit{TESS} spectra. Rotation is substantial in most \dsct stars: single and chemically normal A-type stars show a deficit of slow rotators in the intermediate and late A subtypes \citep{royer_rotational_2007} where \dsct stars are found \citep{murphy_gaia_2019}. Mode identification has long been a major obstacle to interpreting the pulsation spectra of rapidly rotating stars \citep{reese_frequency_2017}. Rotation lifts the degeneracy in azimuthal order $m$, so each mode with degree $\ell \geq 1$ appears as $2\ell + 1$ frequencies. Which components are visible depends on the stellar inclination. A pole-on view suppresses every $m \neq 0$ component, while an equator-on view suppresses those with $\ell + m$ odd parity, which are antisymmetric about the equatorial plane and cancel in the disc integration. In either case the observed spectrum is sparser than the one the star in fact possesses. Rotation also couples modes whose degrees differ by two at fixed $m$, so that neighbouring multiplets overlap and a measured frequency can admit more than one $(n, \ell, m)$ label \citep{dziembowski_effects_1992}. This near-degeneracy becomes important in evolved \dsct models above rotational velocities of roughly 15--20\,km\,s$^{-1}$ \citep{suarez_role_2007}. 

The present paper builds a neural-network emulator and inference pipeline for rotating \dsct pulsation models. As in \citet{scutt_asteroseismology_2023}, the emulator replaces a grid of models with a continuous and rapidly evaluable function. It returns 12 classical and structural quantities, among them $T_{\rm eff}$ and $\log L$, the asteroseismic scalars $\Delta\nu$ and $\varepsilon$, and the modelled $(n, \ell, m)$ frequencies of the p\:modes for $\ell = 0$--$3$, the f\:modes for $\ell = 2$ and $3$, and the g\:modes for $\ell = 1$--$3$ of radial order $n_{\rm pg} = -1$ and $-2$ (Section~\ref{sec:nn}). Training uses the \storm \citep{vanlaer_storm_2025, vanlaer_storm} realisation of the Paper~I grid, in which the Coriolis force is treated directly rather than perturbatively and centrifugal distortion is included to second order. We couple the emulator to a nested sampler through a likelihood that carries its residual errors, including inter-mode correlations, into the posterior (Section~\ref{sec:inference}).

The paper is organised as follows. Section~\ref{sec:methodology} describes the pulsation grids, the neural-network emulator, and the Bayesian inference framework. Section~\ref{sec:validation} presents the emulator residual and synthetic population coverage tests on the grid. Section~\ref{sec:pleiades} applies the framework to \dsct stars in the Pleiades and combines the per-star posteriors into a cluster-level age. We summarise our findings and outline future directions in Section~\ref{sec:conclusions}.

\section{Methodology}
\label{sec:methodology}

Our framework comprises three components: a precomputed grid of stellar pulsation models, a neural-network emulator that learns to interpolate the grid in continuous parameter space, and a Bayesian inference layer that uses the emulator as the forward model when comparing predictions to observed pulsation frequencies. We describe each in turn.

\subsection{Pulsation model grids}\label{sec:grids}

The training data are built on the stellar evolution tracks of the publicly available \dsct pulsation grid presented in Paper~I \citep{gautam_modelling_2025}, with the pulsation frequencies recomputed as described below. We summarise its essential properties here and refer the reader to that paper for full details. The grid covers initial masses, $M$, from 1.4 to 2.5\,M$_\odot$ (in steps of 0.02\,M$_\odot$), initial metal mass fractions, $Z$, from 0.001 to 0.026 (in steps of 0.001), and \emph{initial} equatorial rotation velocities, $v_{\rm eq,0}$, from 0 to 15\,km\,s$^{-1}$ (in steps of 1\,km\,s$^{-1}$)\footnote{These are pre-MS starting values rather than the rotation velocities the models reach. A star spins up as it contracts, so a 1.7\,M$_\odot$, $Z = 0.015$ track begun at $v_{\rm eq,0} = 10$\,km\,s$^{-1}$ arrives at the ZAMS rotating at $v_{\rm eq} = 142.6$\,km\,s$^{-1}$ (Paper~I).}. Stellar evolution tracks were computed with \mesa version r24.03.1 \citep{paxton_modules_2011, paxton_modules_2013, paxton_modules_2015, paxton_modules_2018, paxton_modules_2019, jermyn_modules_2023} from the early pre-main-sequence (pre-MS), through the zero-age main sequence (ZAMS), across the main sequence (MS) to the terminal-age main sequence (TAMS), and through to the end of the post-main-sequence (post-MS) contraction phase. Adiabatic linear pulsation frequencies were computed with \gyre \citep{townsend_gyre_2013, townsend_angular_2018} at each stored model along every track for degrees $\ell = 0$--$3$, including p\:modes for radial orders $n = 1$--$11$, g\:modes for degree $\ell = 1$--$3$ of radial order $n_{\rm pg} = -1$ and $-2$, and f\:modes for $\ell = 2$ and $3$. The original \gyre grid contains approximately $2.5 \times 10^{7}$ stellar pulsation models.

Since \gyre's first-order treatment makes the rotational splitting symmetric in $m$ by construction, we stored each mode compactly in the original \gyre grid. We saved the axisymmetric ($m = 0$) frequency together with a single splitting column $\delta\nu_{n\ell}$; the individual components of a multiplet follow as $\nu_{n\ell m} = \nu_{n\ell 0} + m\,\delta\nu_{n\ell}$. For a more accurate treatment of moderate-rotation effects, we have additionally used the \storm code \citep{vanlaer_storm_2025, vanlaer_storm} to compute the same modes for the same evolutionary tracks. \storm solves the 1D adiabatic pulsation equations with the Coriolis acceleration included directly (rather than perturbatively) for each degree. The centrifugal deformation of the equilibrium structure is then treated perturbatively, using the Chandrasekhar--Milne expansion up to the $P_2$ term, which retains the deformation to $O(\Omega^{2})$, together with the rotational coupling between degrees \citep{vanlaer_storm_2025, mombarg_1d_2025}. The \storm grid stores every computed $(n, \ell, m)$ frequency as it stands, so any asymmetry within a multiplet is retained, and its labels match those used in our mode-identification tables. 

The \storm and \gyre grids share the same evolutionary tracks; what differs is the range of rotation rates over which models are retained. Paper~I restricted its published products to $\Omega/\Omega_{\rm crit} \leq 0.3$, the domain in which a first-order treatment is reliable, whereas here every sub-critical model along a track is kept and the rotation limit is imposed only at inference. At fixed $(M, Z, v_{\rm eq,0})$ the \storm dataset therefore spans a wider range of $\Omega/\Omega_{\rm crit}$, and it comprises approximately $7.1 \times 10^{7}$ models, of which $4.6 \times 10^{7}$ remain after the quality cuts described below. Apart from the retained rotation range and the pulsation treatment itself, the two grids share their input physics and their treatment of rotation in the equilibrium models.

We prepared the \storm grid in two stages. First, at the time of grid building we discarded any track that reaches super-critical rotation ($\Omega/\Omega_{\rm crit} > 1$) at or before the TAMS (usually near or before ZAMS during initial contraction), on the grounds that such a track is unphysical. This removed 790 tracks, or $2.7 \times 10^{6}$ of $7.1 \times 10^{7}$ models. Second, we retained rows with $T_{\rm eff} \geq 6000$\,K, a finite central hydrogen fraction, an age of at least 1\,Myr, and an evolutionary phase at or before the TAMS. This left $4.6 \times 10^{7}$ models.

Beyond the removal of super-critical tracks described above, we impose no rotation cut on the training set, at either the track or the row level. Discarding any track that exceeds a given $\Omega/\Omega_{\rm crit}$ anywhere would remove the moderately rotating main-sequence models the application needs. A star's $\Omega/\Omega_{\rm crit}$ peaks during pre-MS contraction and near the ZAMS, and declines through the remainder of the MS, because the star expands and its surface angular velocity decreases more steeply than its critical rate. The MS models at the rotation rates of our targets therefore lie on tracks that exceed any such limit earlier in their evolution. A row-level cut would keep those tracks but excise segments from within them, leaving gaps in age along tracks that the grid otherwise samples continuously, which the emulator would then have to interpolate across. We therefore train on the full sub-critical grid and impose the rotation limit only at inference (Section~\ref{sec:inference_priors}).

We adopt $\Omega/\Omega_{\rm crit} \leq 0.5$ as the inference domain. This is less conservative than the $\Omega/\Omega_{\rm crit} \lesssim 0.3$ limit used for the first-order \gyre products in Paper~I, because \storm treats the Coriolis force directly and includes the second-order centrifugal deformation in the pulsation calculation, where \gyre retains only the first-order splitting. The only published test of \storm against two-dimensional \textsc{ester} structures \citep{rieutordAlgorithmComputing2D2016a} and \textsc{top} pulsation calculations \citep{reeseOscillations2DESTER2021} used $\beta$\,Cephei models and reached $\Omega/\Omega_{\rm crit,Kep} = 0.2$ \citep{mombarg_1d_2025}. No equivalent benchmark exists across the \dsct domain.

The perturbative treatment of rotation sets an upper limit on where our models can be trusted, and that limit is tighter for g\:modes than for p\:modes. \citet{ballot_gravity_2010} find second-order perturbative methods suited to low-order, low-degree g\:modes up to equatorial velocities near $100$\,km\,s$^{-1}$ in $\gamma$\,Dor models. The grid holds g\:modes only at $\ell = 1$--$3$ and $n_{\rm pg} = -1$ and $-2$, which is the low-order, low-degree regime to which that bound applies.

\subsection{Neural-network emulator}\label{sec:nn}

\subsubsection{Problem definition}\label{sec:nn_problem_fo_resmlp}

The emulator replaces the stellar-evolution and pulsation calculations during inference with a deterministic regression function
\begin{equation}
    \hat{\mathbf{y}} = f_{\mathbf{w}}(\boldsymbol\theta),
\end{equation}
parameterised by the network weights $\mathbf{w}$. Throughout this paper, a hat marks an emulator prediction. The grid value it approximates, $\mathbf{y}$, carries no hat, and a measured value carries the subscript ``obs''. The four-dimensional input vector is
\begin{equation}
    \boldsymbol\theta = (M,\,Z,\,\tau,\,v_{\rm eq}),
\end{equation}
where $M$ is the stellar mass, $Z$ is the initial metal mass fraction, $\tau$ is the absolute stellar age in Myr, and $v_{\rm eq}$ is the surface equatorial rotation velocity at that age\footnote{$v_{\rm eq}$ is the value the model has reached at that age rather than the initial value $v_{\rm eq,0}$ that labels the grid axis (Section~\ref{sec:grids})}. Using age directly keeps the sampling coordinates physically interpretable and avoids the post-sampling conversion required when an evolutionary coordinate such as the central hydrogen abundance is used \citep[cf.][]{dotter_2016, mombarg_constraining_2021}. Central hydrogen also provides little age resolution in young stars because it remains close to its initial value until core hydrogen burning is well established. The broad dynamic range in age is handled by the input transformation described in Section~\ref{sec:nn_preproc_fo_resmlp}.

The output vector contains 12 classical and structural stellar quantities: effective temperature $T_{\rm eff}$, $\log L$, $\log R$, $\log g$, mean density $\bar{\rho}$, central hydrogen fraction $X_{\rm c}$, equatorial radius $R_{\rm eq}$, polar radius $R_{\rm pol}$, critical angular velocity $\Omega_{\rm crit}$, the surface-to-critical rotation ratio $\Omega/\Omega_{\rm crit}$, and the pp-chain and CNO-cycle luminosities. The network also predicts the asteroseismic scalars $\Delta\nu$ and $\varepsilon$, together with 218 individual \storm frequencies: 176 p\:modes, 12 f\:modes, and 30 g\:modes. The complete output vector therefore contains 232 quantities per stellar model.

\subsubsection{Preprocessing}\label{sec:nn_preproc_fo_resmlp}

We trained the emulator on the \storm grid after the track removals, label masking and row cuts described in Section~\ref{sec:grids}. We used a deterministic hybrid row-and-track split to divide the data into 80\,per\,cent training, 15\,per\,cent validation, and 5\,per\,cent testing. Approximately 1\,per\,cent of all evolutionary tracks were assigned to testing and 3\,per\,cent to validation, with all models along each selected track retained together. Rows sampled from the remaining tracks contribute nominally a further 4\,per\,cent of the full dataset to testing and 12\,per\,cent to validation.

The two held-out components test different aspects of the emulator. Rows sampled from tracks represented during training test local interpolation along an evolutionary track. Complete held-out tracks instead test generalisation to grid tracks with unseen combinations of mass, composition, and initial rotation.

The four inputs ($M$, $Z$, $v_{\rm eq}$, $\tau$) have different numerical ranges, so we transformed and standardised each one before training. We used a separate Yeo--Johnson transformation for each input \citep{yeo_johnson_2000}, implemented with \textsc{scikit-learn} \citep{pedregosa_scikit_2011}. This reduces skewness and places the inputs on comparable scales. Because the Yeo--Johnson transformation is defined at zero, it also handles the $v_{\rm eq}=0$ models without needing an arbitrary offset.

Each of the 176 p\:mode frequencies receives a separate Yeo--Johnson transformation. Each of the 30 g\:mode frequencies is centred on its median and divided by its interquartile range, and the same scaling is applied separately to $\bar{\rho}$, $X_{\rm c}$, and $\Omega/\Omega_{\rm crit}$. This robust scaling is particularly useful for $X_{\rm c}$, whose training-set distribution has a long lower tail. The f\:modes, $\Delta\nu$, $\varepsilon$, and remaining stellar outputs are each centred on their mean and divided by their standard deviation. All centring and scaling statistics are calculated from finite training-set values.

Acoustic frequencies are often normalised by $\sqrt{GM/R^3}$, which is proportional to $\sqrt{\bar{\rho}}$, to make them dimensionless and reduce their variation along evolutionary tracks. We do not apply this scaling before the transformations described above because inverting that scaling multiplies by the network's own predicted density, so an error in a single output would propagate into every frequency. We trained a variant with the scaling applied and found no improvement in frequency accuracy. All transformations were fitted using the training partition alone, so neither the validation nor test targets influence the preprocessing.

\subsubsection{Architecture}\label{sec:nn_arch_fo_resmlp}

\begin{figure*}
    \centering
    \includegraphics[width=\textwidth]{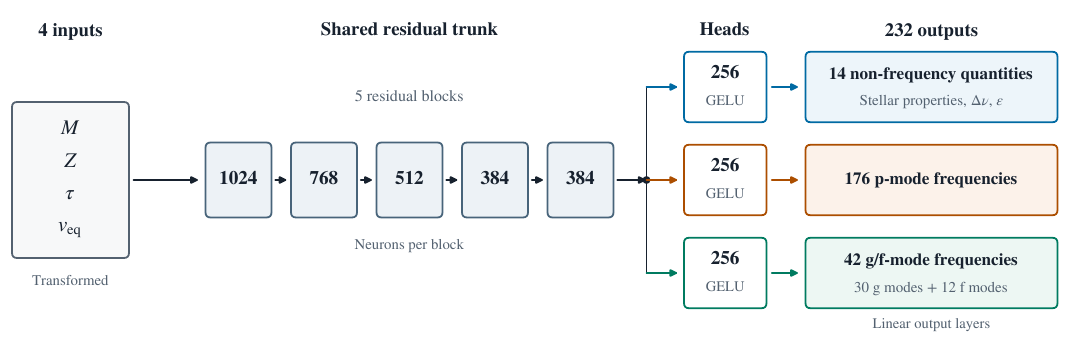}
    \caption{Architecture of the residual multi-head emulator. Four transformed stellar inputs pass through five shared residual blocks; the numbers in the boxes give their neuron counts. Each block includes an affine transformation, a GELU activation and a residual shortcut, with a learned projection when the input and output dimensions differ. The final 384-neuron representation feeds three specialised heads, each with a 256-neuron GELU hidden layer and a linear output layer. Together they predict 14 non-frequency quantities, 176 p\:mode frequencies and 42 g/f\:mode frequencies directly. The network has 3.27 million trainable parameters.}
    \label{fig:nn_architecture}
\end{figure*}

To predict the stellar properties and pulsation frequencies within a single emulator, we use a residual multilayer perceptron with a shared trunk and three specialised output heads. This branching design is inspired by \textsc{Pitchfork} \citep{scutt_pitchfork_2026}, which uses shared layers to learn information common to the classical and asteroseismic outputs before separating into specialised branches. We adapt this principle to the \storm target set by using separate heads for the non-frequency quantities, the p\:modes, and the combined g/f\:mode block. Whereas \textsc{Pitchfork} represents its pulsation frequencies through principal component analysis (PCA), a dimensionality reduction technique, and reconstructs them from the PCA coefficients, our emulator predicts the complete 232-dimensional output vector directly. Figure~\ref{fig:nn_architecture} summarises the resulting architecture.

The shared trunk maps the four transformed stellar inputs to a representation used by all three output heads. It contains five residual blocks, of 1024, 768, 512, 384 and 384 neurons. Each of the five trunk blocks multiplies its input by a learned weight matrix and adds a learned bias, $\mathbf{z}=\mathbf{W}\mathbf{x}+\mathbf{b}$. This is followed by a Gaussian Error Linear Unit \citep[GELU;][]{hendrycks_gelu_2016}, which applies the smooth nonlinear function $x\,\Phi(x)$ to each neuron, where $\Phi$ is the standard normal cumulative distribution function. Without such a nonlinearity, a stack of affine transformations would collapse to a single affine transformation, and the emulator could only represent quantities that vary linearly with $M$, $Z$, $\tau$ and $v_{\rm eq}$. The block input is then added to the activated output through a residual connection. When a block's input and output contain different numbers of neurons, a learned linear projection maps the residual to the required size.

Each of the three heads takes the 384-neuron output of the final trunk block through one hidden layer of 256 neurons with a GELU activation, followed by a linear output layer: 14 neurons for the non-frequency quantities, 176 for the p\:modes, and 42 for the combined g/f\:mode block. We apply no dropout, and instead rely on weight decay and early stopping (Section~\ref{sec:nn_loss_fo_resmlp}). Counting the learned projections in the residual connections, the network has 3.27 million trainable parameters.

\subsubsection{Loss and optimisation}\label{sec:nn_loss_fo_resmlp}

Training minimises a masked mean absolute error (MMAE) in transformed output space. We calculate a separate loss for each output head: non-frequency ($\mathrm{c}$), p\:mode ($p$), and combined g/f\:mode ($gf$). Within each minibatch, let $\mathcal{V}_h$ be the set of index pairs $(i,k)$ for which model $i$ has a finite target value for output $k$ of head $h$. The loss for head $h$ is
\begin{equation}
    \mathcal{L}_h =
    \frac{1}{|\mathcal{V}_h|}
    \sum_{(i,k)\in\mathcal{V}_h}
    \left|
    \tilde{y}_{i,k}^{\rm pred}
    -
    \tilde{y}_{i,k}
    \right|,
\end{equation}
where tildes denote transformed outputs and $|\mathcal{V}_h|$ is the number of finite target entries for head $h$ in the minibatch. We combine the three head losses using a weighted mean:
\begin{equation}
    \mathcal{L}_{\rm total}
    =
    \frac{
        3\mathcal{L}_{\rm c}
        +7\mathcal{L}_{p}
        +5\mathcal{L}_{gf}
    }{3+7+5}.
    \label{eq:loss_fo_resmlp}
\end{equation}
The weights emphasise the p\:mode spectrum, which holds most of the constraining power at inference, while retaining substantial contributions from the g/f\:mode and non-frequency heads. They were chosen by hand from a set of emulators trained with different weights rather than by a systematic search. We compute the mean within each group before applying them to prevent the number of columns in a group from determining its contribution by itself, which would otherwise leave the 14 non-frequency outputs negligible beside the 176 p\:modes. The element-wise mask allows a stellar model with a missing pulsation mode frequency to contribute through every output that is available. We minimise absolute rather than squared error so that unusually large residuals do not dominate the training gradient. Under squared error those cells would contribute in proportion to the square of their size. Section~\ref{sec:residuals} shows that the largest residuals are concentrated in the final pre-main-sequence contraction, outside the evolutionary window of the Pleiades stars.

\begin{table}
    \centering
    \caption{Training settings of the emulator.}
    \label{tab:nn_training}
    \begin{tabular}{ll}
        \hline
        Setting & Value\\
        \hline
        Loss & masked mean absolute error, Eq.~\eqref{eq:loss_fo_resmlp}\\
        Optimiser & \textsc{AdamW}\\
        Initial LR& $5\times10^{-4}$\\
        LR reduction factor & 0.7 after 5 epochs without improvement\\
        Minimum LR & $10^{-6}$\\
        Weight decay & $10^{-5}$\\
        Minibatch size & 4096 models\\
        Maximum gradient norm & 1.0\\
        Precision & 32-bit\\
        Training time & 72\,h on one NVIDIA V100S GPU\\
        Adopted weights & epoch 942, validation loss $6.23\times10^{-4}$\\
        \hline
    \end{tabular}
\end{table}

We trained the network from scratch with \textsc{AdamW} \citep{loshchilov_decoupled_2019}, a gradient-descent optimiser that adapts the learning rate (LR) of each weight individually, using the settings in Table~\ref{tab:nn_training}. \textsc{AdamW} also applies weight decay, which penalises large weights and so discourages the network from fitting noise in the training set. We updated the weights after each minibatch of training models rather than after each complete pass through the training set (an epoch), and reduced the learning rate whenever the validation loss stopped improving. We also clip the global gradient norm at 1.0, which limits how far a single minibatch can move the weights. Training stops once the validation loss has not improved for 150 epochs, or after 5000 epochs, whichever comes first. Stopping on the validation loss rather than the training loss guards against overfitting.  

We retained the weights from the epoch of lowest validation loss, which for this run was epoch 942, where Eq.~\eqref{eq:loss_fo_resmlp} reached $6.23\times10^{-4}$. We use this trained emulator for all results that follow.

With the released code (see the \hyperref[sec:data_availability]{Data Availability} section), the trained emulator returns all 232 outputs for one set of inputs in about 1.4\,ms on one core of a laptop CPU (Apple M2 Pro). Most of this time is a fixed cost per call. Predicting $10^{4}$ models in one vectorised call reduces the cost to about 14\,$\mu$s per model, so $10^{6}$ predictions take about 14\,s. A complete fit of one star at the settings of Section~\ref{sec:inference_sampler} needs about $10^{5}$ likelihood evaluations and takes 1.2--1.5\,min on the same core.

\subsection{Bayesian inference framework}\label{sec:inference}

\subsubsection{Forward model}\label{sec:inference_forward}

At inference time, the trained emulator maps the four input parameters to predicted observables. We apply the input transformation of Section~\ref{sec:nn_preproc_fo_resmlp}, run the network, and invert the output transformations to return physical units. The priors of Section~\ref{sec:inference_priors} remain inside the training range of each input parameter.

\begin{figure*}
    \centering
    \includegraphics[width=\textwidth]{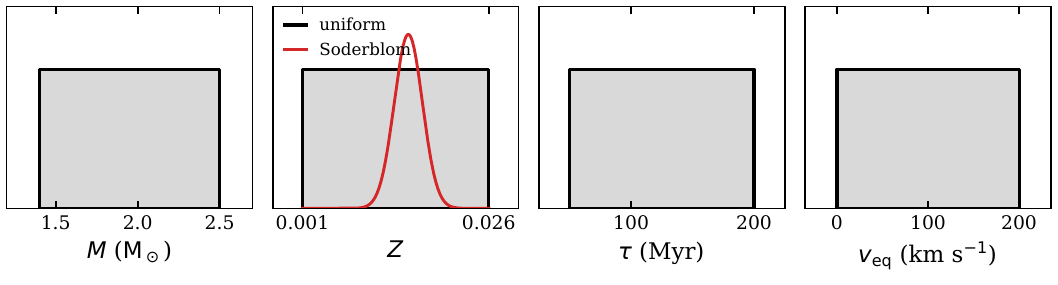}
    \caption{Priors used for the Pleiades inference. Mass, age and equatorial velocity have uniform priors over the displayed ranges. For metallicity, the red curve shows the truncated Gaussian distribution of \citet{soderblom_metallicity_2009} adopted for the Pleiades application; the black curve shows the uniform prior over the emulator grid used in the sensitivity test of Appendix~\ref{app:metallicity_prior}. The age range shown is the Pleiades specific window.}
    \label{fig:inference_priors}
\end{figure*}

We characterise the emulator's residuals on the held-out test set by a standard deviation for each output:
\begin{equation}
    \hat{\sigma}_{{\rm model},k} = \sqrt{\frac{1}{N_{\rm test}} \sum_{i=1}^{N_{\rm test}} \left(d_{i,k} - \bar{d}_{k}\right)^{2}},
    \label{eq:sigma_model}
\end{equation}
where $d_{i,k}=\hat y_{i,k}-y_{i,k}$ is the prediction minus the grid-truth value of output $k$ at test model $i$, and $\bar d_k$ is its test-set mean. The vector $\hat{\boldsymbol\sigma}_{\rm model}$ sets the diagonal emulator-error scale in the likelihood (Section~\ref{sec:inference_likelihood}).

\subsubsection{Priors}\label{sec:inference_priors}

We place broad, physically motivated priors on the four input parameters (Fig.~\ref{fig:inference_priors}). For mass we use a uniform prior on $[1.4, 2.5]\,\mathrm{M}_\odot$ matching the grid coverage. For the Pleiades application, we adopt a metallicity prior based on the spectroscopic measurement of \citet{soderblom_metallicity_2009}: $Z \sim \mathcal{N}(0.0152,\,0.0019)$, truncated at the grid boundaries. For the equatorial rotation velocity we use a uniform prior on $[0,\,200]$\,km\,s$^{-1}$, comfortably spanning the rotation velocities of our slow- to moderately-rotating targets. For the age dimension we place a uniform prior spanning a cluster appropriate window of $\tau \in [50, 200]$\,Myr for the Pleiades targets. The separate $\Omega/\Omega_{\rm crit}$ constraint is described below.

We enforce one hard constraint: any sample whose predicted ratio $\Omega / \Omega_{\rm crit}$ exceeds $0.5$ is rejected with $\ln \mathcal{L} = -\infty$. Since no rotation limit was applied to the training data (Section~\ref{sec:grids}), this is where the rotation limit enters. The constraint is implemented inside the likelihood, so the sampler does not visit these regions of parameter space. This constraint never triggers for the targets studied here because the posteriors of our Pleiades targets, and every target of the population tests of Section~\ref{sec:hare_hounds}, remain far below the threshold; the largest $\Omega / \Omega_{\rm crit}$ ratio sampled for any of the five Pleiades stars is $0.23$. 

\subsubsection{Likelihood}\label{sec:inference_likelihood}

For each star we compare the emulator's predicted frequencies $\hat{\boldsymbol\nu}(\boldsymbol\theta)$ and predicted effective temperature $\hat T_{\rm eff}(\boldsymbol\theta)$ to the corresponding observations through a \emph{multivariate} Gaussian likelihood. Writing the frequency residual vector as $\mathbf{r}(\boldsymbol\theta) = \boldsymbol\nu_{\rm obs} - \hat{\boldsymbol\nu}(\boldsymbol\theta)$, the frequency term is
\begin{equation}
    \ln \mathcal{L}_\nu(\boldsymbol\theta) = -\frac{1}{2}\left[ \mathbf{r}^{\!\top} \mathrm{C}_\nu^{-1}\, \mathbf{r} + \ln\!\left|\mathrm{C}_\nu\right| + n_\nu \ln(2\pi) \right],
    \label{eq:likelihood}
\end{equation}
where $\mathrm{C}_\nu$ is the $n_\nu \times n_\nu$ frequency covariance matrix defined below and $n_\nu$ is the number of fitted modes. The first term is a $\chi^2$ that accounts for correlations between mode errors. Since $\mathrm{C}_\nu$ is fixed during each fit, we factorise it once as $\mathrm{C}_\nu = \mathrm{L}\mathrm{L}^{\!\top}$, with $\mathrm{L}$ lower triangular (a Cholesky decomposition), avoiding explicit matrix inversion \citep[Appendix~A.4]{rasmussen_williams_2006}. Each likelihood evaluation then uses only a single triangular solve, which is fast and numerically stable. Effective temperature is the only classical observable included and contributes a separate Gaussian term, so the total log-likelihood is
\begin{equation}
    \ln \mathcal{L}(\boldsymbol\theta) = \ln \mathcal{L}_\nu(\boldsymbol\theta) + \ln \mathcal{L}_{T_{\rm eff}}(\boldsymbol\theta),
    \label{eq:likelihood_total}
\end{equation}
with
\begin{equation}
\begin{split}
    \ln \mathcal{L}_{T_{\rm eff}}(\boldsymbol\theta) &= -\frac{1}{2}\left[ \frac{\left(T_{\rm eff,obs} - \hat T_{\rm eff}(\boldsymbol\theta)\right)^{2}}{\sigma_{c}^{2}} + \ln \sigma_{c}^{2} + \ln(2\pi) \right], \\
    \sigma_{c}^{2} &= \sigma_{T_{\rm eff,\rm obs}}^{2} + \sigma_{T_{\rm eff,\rm emu}}^{2},
\end{split}
    \label{eq:likelihood_teff}
\end{equation}
where $\sigma_{T_{\rm eff,\rm obs}}$ is the observational uncertainty and $\sigma_{T_{\rm eff,\rm emu}}$ is the emulator residual standard deviation for $T_{\rm eff}$ over the full test set, $\hat{\sigma}_{\rm model}$ in Eq.~\eqref{eq:sigma_model}.

\begin{figure}
    \centering
    \includegraphics[width=\columnwidth]{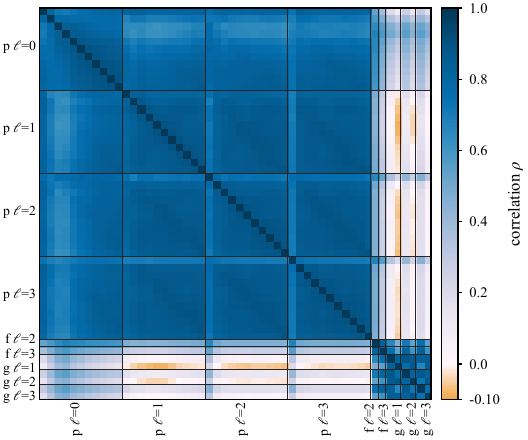}
    \caption{Correlation between the emulator's frequency residuals for pairs of modes, conditioned on the Pleiades prior of Section~\ref{sec:inference_priors}. This is $\mathrm{C}_{\rm emu}$ of Eq.~\eqref{eq:cov_emu} as the Pleiades \dsct fits used it, normalised by its diagonal. Each row and each column is one predicted frequency, and the cell where they meet is the correlation between the emulator's error on those two modes. A positive value means that when the emulator predicts one of the two modes too high it tends to predict the other too high as well, a negative value that it tends to predict the other too low, and zero that the error on one says nothing about the error on the other. Modes are grouped into blocks of common family and degree, which the axis labels name, and within each block the rows run in radial order, so the individual orders are not labelled. 
    }
    \label{fig:emu_correlation}
\end{figure}

The matrix $\mathrm{C}_\nu$ is the covariance of the frequency \emph{errors}, not the covariance of frequencies across the grid (which would encode astrophysical variation rather than error). We build it as the sum of two physically distinct contributions,
\begin{equation}
    \mathrm{C}_\nu = \mathrm{C}_{\rm obs} + \mathrm{C}_{\rm emu},
    \label{eq:cov_terms}
\end{equation}
defined as follows.
\begin{description}
    \item[Observational, $\mathrm{C}_{\rm obs}$.] A diagonal matrix $\mathrm{diag}(\sigma_{{\rm obs},i}^2)$ of the per-frequency uncertainties from the prewhitening. Each $\sigma_{{\rm obs},i}$ is the formal least-squares uncertainty of a coherent sinusoid in white noise \citep{montgomery_derivation_1999}, evaluated in the amplitude-spectrum form of \citet{kjeldsenKeplerCoRoTMOST2011} from the noise level of the residual spectrum after the final prewhitening step, and is inversely proportional to the mode amplitude and the  duration $T$ of the light curve. These are lower limits, since correlated noise and unresolved neighbouring peaks are not accounted for, but for the fitted modes they lie between $10^{-6}$ and $2\times10^{-3}$\,d$^{-1}$ (Appendix~\ref{app:modeids}), at least an order of magnitude below the emulator term, so $\mathrm{C}_{\rm obs}$ adds less than one per cent to any diagonal element of $\mathrm{C}_\nu$. We discuss the limitation of not modelling correlations among observational frequency errors introduced by prewhitening a single time series in Section~\ref{sec:conclusions}.

    \item[Emulator, $\mathrm{C}_{\rm emu}$.] We combine correlations estimated within the prior region with error scales from the full test split. For the Pleiades prior of Section~\ref{sec:inference_priors}, we select $1.5\times10^{4}$ of the $2.3\times10^{6}$ test models and weight them by the prior density. With weights normalised so that $\sum_i w_i=1$, the effective sample size is $(\sum_i w_i^2)^{-1}=4.1\times10^{3}$.

    For selected test model $i$ and mode $k$, the emulator residual is $d_{ik}=\hat\nu_k(\boldsymbol\theta_i)-\nu_{{\rm true},k}(\boldsymbol\theta_i)$, with weighted mean $\bar d_{\pi,k}=\sum_i w_i d_{ik}$. We estimate the raw covariance as
    \begin{equation}
        \mathrm{C}_{{\rm raw},jk}
        =
        \frac{
            \sum_i w_i(d_{ij}-\bar d_{\pi,j})(d_{ik}-\bar d_{\pi,k})
        }{
            1-\sum_i w_i^2
        },
        \label{eq:cov_raw}
    \end{equation}
    where the sums run over the selected test models. Each element has units of d$^{-2}$. The corresponding correlation matrix,
    \begin{equation}
        \mathrm{R}_{\pi,jk}=\mathrm{C}_{{\rm raw},jk}/\sqrt{\mathrm{C}_{{\rm raw},jj}\mathrm{C}_{{\rm raw},kk}},
        \label{eq:corr}
    \end{equation}
    is combined with the full test-set standard deviations from Eq.~\eqref{eq:sigma_model} as $\mathrm{D}=\mathrm{diag}(\hat{\sigma}_{{\rm model},k})$. We then construct
    \begin{equation}
        \mathrm{C}_{\rm emu}
        = \mathrm{D}\big[(1-\alpha)\mathrm{R}_{\pi}+\alpha\mathrm{I}\big]\mathrm{D}
        + \eta_{\rm num}\mathrm{I},
        \label{eq:cov_emu}
    \end{equation}
    where $\mathrm{I}$ is the identity matrix. We fix $\alpha=0.1$, shrinking the off-diagonal correlations towards zero by 10\,per\,cent. We also add $\eta_{\rm num}=10^{-8}\times\mathrm{median}_k(\hat{\sigma}_{{\rm model},k}^{\,2})$ to each diagonal element to improve the numerical stability of the Cholesky factorisation.

    Figure~\ref{fig:emu_correlation} shows the resulting correlations. We compute $\mathrm{C}_{\rm emu}$ once at setup; it supplies the only off-diagonal terms in $\mathrm{C}_\nu$.
\end{description}

We omit luminosity from the likelihood because published estimates often use temperature-dependent bolometric corrections and may therefore be correlated with the adopted $T_{\rm eff}$ constraint. More generally, uncertain extinction in dusty regions and additional extinction or emission from circumstellar discs can make stellar luminosities difficult to determine reliably.

Each stellar posterior is conditional on a fixed table of observed frequencies, uncertainties, and assigned $(n,\ell,m)$ labels; Section~\ref{sec:modeid} describes how these tables are constructed and assessed.

\subsubsection{Sampler}\label{sec:inference_sampler}

We sample the posterior with the reactive nested sampling implementation \ultranest \citep{buchner_ultranest_2021}. Nested sampling explores the prior volume in nested likelihood shells and can resolve multimodal posteriors. We start with 1000 live points, request an evidence tolerance of \texttt{dlogz}$=0.1$ and a minimum effective posterior sample size of 1000, and cap each run at $10^{7}$ likelihood evaluations. The sampler returns samples on $(M,Z,\tau,v_{\rm eq})$ and the log-evidence $\ln\mathcal{Z}$, which we report as a sampling diagnostic. Appendix~\ref{app:convergence} distinguishes the final logged error-budget estimates from the returned evidence uncertainties and gives the achieved sample sizes and evaluation counts. As a diagnostic of the frequency fit, we report $\chi^2_{\nu,\rm freq} = \mathbf{r}^{\!\top}\mathrm{C}_\nu^{-1}\mathbf{r}/n_\nu$, normalised by the number of fitted modes and evaluated at the posterior median.

\section{Validation: emulator residuals and population coverage tests}
\label{sec:validation}

Before applying the framework to observations, we assess the emulator's prediction errors against held-out grid models (Section~\ref{sec:residuals}) and test parameter recovery and credible-interval coverage with synthetic targets (Section~\ref{sec:hare_hounds}). These targets are generated either by the emulator itself, to check the statistical implementation, or by held-out grid models, to assess recovery in the presence of interpolation error.

\subsection{Emulator residuals}\label{sec:residuals}

\begin{figure*}
    \centering
    \includegraphics[width=\textwidth]{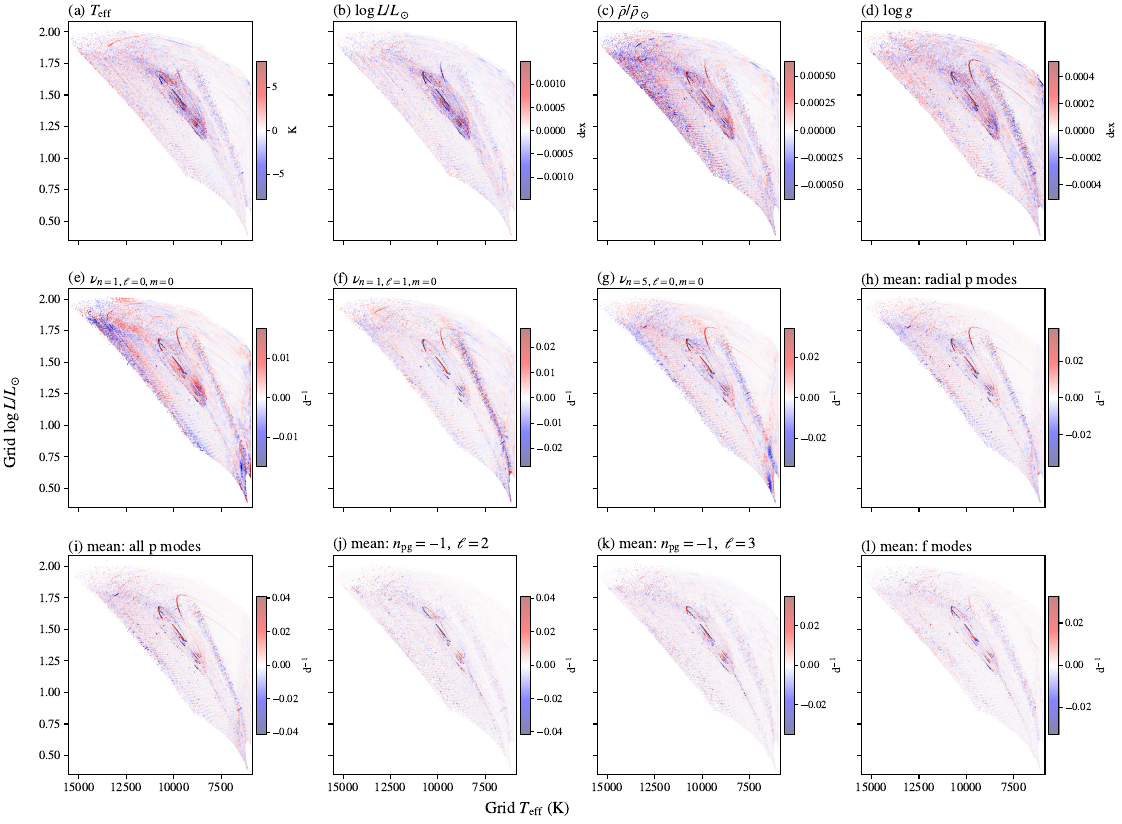}
    \caption{The emulator residuals across the Hertzsprung--Russell diagram for the approximately $2.3\times10^{6}$ models in the held-out test split. Each point is placed at its grid-truth $(T_{\rm eff},\log L/L_\odot)$ and coloured by the signed residual $\Delta y=y_{\rm NN}-y_{\rm grid}$ in physical units. Panels (a)--(d) show the residuals in $T_{\rm eff}$, $\log L$, mean density, and $\log g$; panels (e)--(g) show three representative individual p\:modes; and panels (h)--(l) show the per-model mean signed frequency residual over all available radial p\:modes, all p\:modes, the $n_{\rm pg}=-1$, $\ell=2$ modes, the $n_{\rm pg}=-1$, $\ell=3$ modes, and the f\:modes, respectively. Each panel has an independent symmetric colour scale clipped at the 99th percentile of $|\Delta y|$ so that the structure of the central residual distribution remains visible. Values outside the displayed range are saturated.}
    \label{fig:nn_validation}
\end{figure*}

We evaluated the trained network on every model in the held-out test split and tabulated per-output residuals between the predicted and grid truth values. The test split contains $N_{\rm test} \approx 2.3\times10^{6}$ models, large enough to characterise residuals across the full input parameter space. Figure~\ref{fig:nn_validation} shows the signed residuals in physical units across the Hertzsprung--Russell diagram, both for individual outputs and for averages over relevant mode families.

The p\:modes and f\:modes have mean absolute errors (MAEs) of $4.1\times10^{-3}$ and $2.4\times10^{-3}$\,d$^{-1}$, respectively. These exceed the formal frequency uncertainties of every fitted mode, typically by nearly two orders of magnitude, so the emulator error dominates the frequency term in the likelihood. The p\:mode residual grows monotonically with radial order in absolute terms, from $2.7\times10^{-3}$\,d$^{-1}$ at $n = 1$ to $6.4\times10^{-3}$\,d$^{-1}$ at $n = 11$, but it grows more slowly than the frequency itself, so the fractional error falls with radial order. The mean fractional error over the p\:modes is $8.5\times10^{-5}$.

\begin{figure*}
    \centering
    \includegraphics[width=0.49\textwidth]{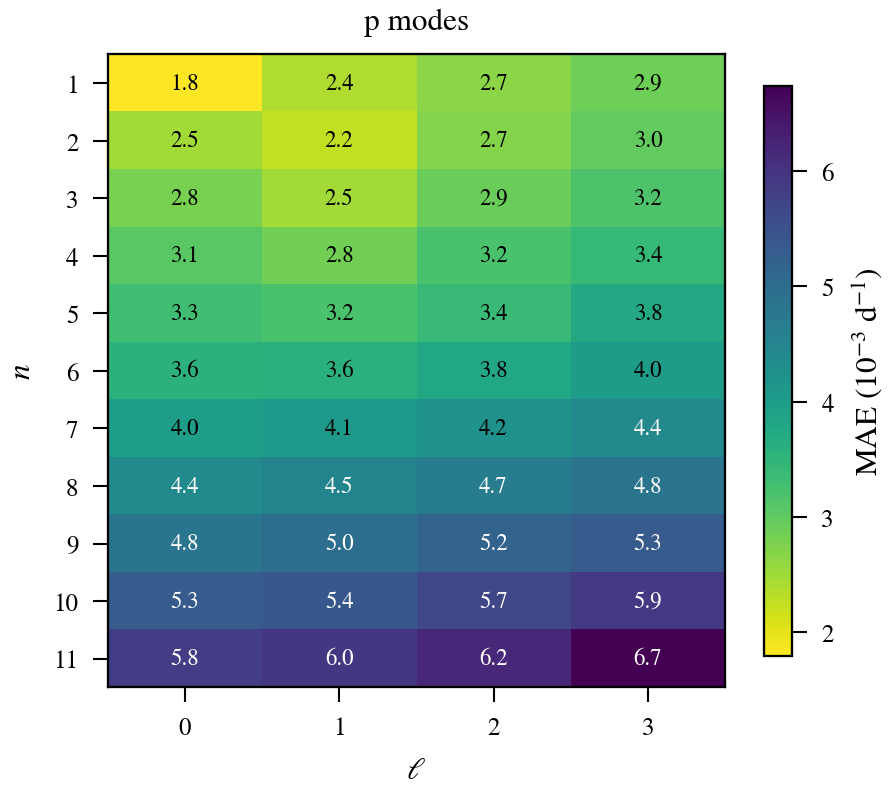}
    \includegraphics[width=0.49\textwidth]{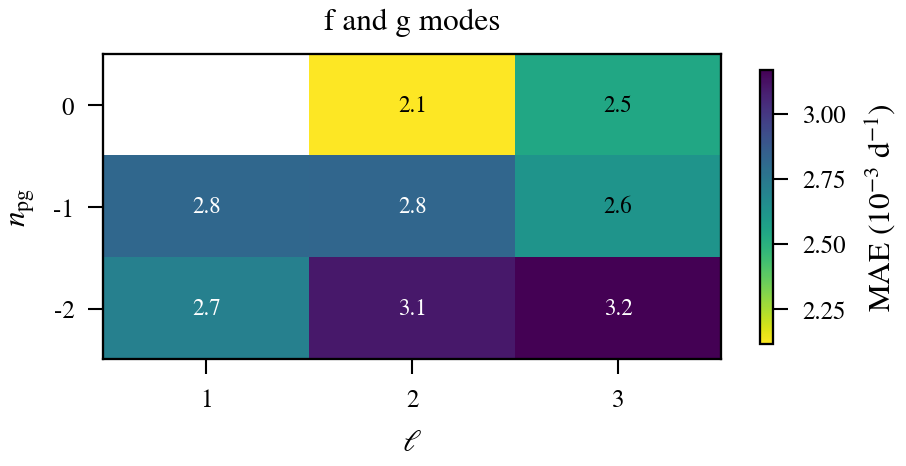}
    \caption{Per-mode mean absolute error heatmaps for the emulator on the held-out test split, in physical units (d$^{-1}$). \emph{Left:} p\:modes, indexed by radial order $n$ and degree $\ell$. \emph{Right:} f and g\:modes, indexed by $n_{\rm pg}$; the $n_{\rm pg}=0$ row holds the f\:modes, which exist only for $\ell \geq 2$, and negative orders are g\:modes (radial order increasing downward in frequency).}
    \label{fig:mae_heatmaps}
\end{figure*}

The g\:modes are recovered at the same level, with a MAE of $2.9\times10^{-3}$\,d$^{-1}$ over the 30 retained columns. As the heatmap shows (Fig.~\ref{fig:mae_heatmaps}), the per-degree means at $n_{\rm pg} = -1$ and $-2$ all lie between $2.6\times10^{-3}$ and $3.2\times10^{-3}$\,d$^{-1}$, with no strong dependence on degree or radial order.

Classical-observable residuals are smaller than the observational uncertainties adopted in this work. The mean absolute $T_{\rm eff}$ residual is $0.8$\,K, with a root-mean-square of $11.3$\,K; $\log L$ and mean density have mean absolute residuals of $1.5\times10^{-4}$ and $6.7\times10^{-5}$, respectively.

The models with large residuals form an identifiable population rather than a scatter across the grid. Models aged 3--10\,Myr contribute over 99\,per\,cent of the test-set mean-square p\:mode residual: the per-model mean p\:mode residual has a root mean square of $7.1\times10^{-2}$\,d$^{-1}$ there, against $5$--$6\times10^{-3}$\,d$^{-1}$ at ages of 10--100\,Myr. The most extreme cases are contracting models of $\gtrsim 2.1\,M_\odot$ at ages of 5.5--8\,Myr, with central hydrogen still unburnt, moderately rapid rotation, and compositions on the metal-rich side of the grid. The emulation difficulty arises from the rapid structural evolution of that phase rather than proximity to a grid edge. Excluding models younger than 10\,Myr leaves thin-tailed residuals for the classical observables and p\:modes (root-mean-square $T_{\rm eff}$ residual $0.7$\,K, per-model mean p\:mode residual $7.2\times10^{-3}$\,d$^{-1}$), and the remaining large residuals are confined to the $n_{\rm pg}=-1$, f\:mode, and low-order $\ell=1$ columns in a narrow band along the evolved main sequence. These are the same mode families, in the same phase of evolution, in which the recovery tests of Section~\ref{sec:hare_hounds} and Appendix~\ref{app:recovery_failures} find the emulator error concentrated by the avoided crossings between the low-order g, f, and p\:modes. These are localised in the fraction of the initial central hydrogen remaining at $X_{\rm c}/X_0 \approx 0.77$--$0.91$.

The residuals also define the per-output emulator standard deviation $\hat{\sigma}_{\rm model}$ of Eq.~\eqref{eq:sigma_model}, which we use to characterise the emulator's accuracy output by output. It is an order of magnitude larger than the mean absolute error of the same outputs: the median $\hat{\sigma}_{\rm model}$ over the p\:modes is $5.7\times10^{-2}$\,d$^{-1}$ against a MAE of $4.1\times10^{-3}$\,d$^{-1}$, and the median over the f and $n_{\rm pg}=-1$ modes is $2.7\times10^{-2}$\,d$^{-1}$. The two differ because the residual distribution is heavy-tailed. The small minority identified above, almost all in the final pre-main-sequence contraction, carry much larger residuals and dominate a variance computed over the whole test set. This tail is the population that a squared-error loss would have weighted most strongly, which is why we train on absolute error (Section~\ref{sec:nn_loss_fo_resmlp}). The largest single value among the frequencies used in the five adopted fits is $6.7\times10^{-2}$\,d$^{-1}$, on the $n=8$, $\ell=1$ column. For $T_{\rm eff}$, $\hat{\sigma}_{\rm model}$ is $11.3$\,K.

Figure~\ref{fig:emu_correlation} shows the residual correlations used in Eq.~\eqref{eq:cov_emu}: the emulator's p\:mode errors move almost as one, while its p and g\:mode errors are nearly uncorrelated.

\subsection{Population recovery and coverage tests}\label{sec:hare_hounds}

We test whether the full inference pipeline recovers parameters and calibrated credible intervals from synthetic mode lists resembling those of the Pleiades stars. All tests use the likelihood and covariance construction of Section~\ref{sec:inference_likelihood}. Three populations differ in their source of truth and injected noise:
\begin{description}
    \item[Emus.] The truth observables are generated by the emulator itself, $\hat{\boldsymbol\nu}(\boldsymbol\theta_{\rm true})$. Because the forward model and the data-generating model are the same, the frequency residual is drawn from $\mathrm{C}_\nu$ and the effective-temperature residual from the scalar variance $\sigma_c^2$. This isolates the statistical machinery (likelihood, covariance and sampler) under an exactly specified data-generating model.
    \item[Hares.] The truth observables, $T_{\rm eff}$ and the selected \storm mode frequencies, are taken from grid models in the held-out test split. No noise is injected, so this population isolates the pure emulator-versus-grid systematic at each target.
    \item[Noisy hares.] Same as for the hares, but with measurement noise added. Each frequency receives an independent Gaussian perturbation scaled by the observational uncertainty assigned to that mode (Section~\ref{sec:hh_design}), and $T_{\rm eff}$ receives a Gaussian perturbation of $250$\,K.
\end{description}

\subsubsection{Campaign design}\label{sec:hh_design}

Calibration can vary with evolutionary phase, so we stratified the campaign into age panels rather than pooling all targets. Age panels let each fit use an informative age prior like the Pleiades analysis while sampling different evolutionary phases. Section~\ref{sec:hh_global} also examines the results directly as a function of evolutionary phase.

We selected 100 held-out targets in each of four age panels: $1$–$50$, $50$–$200$, $200$–$500$ and $500$–$1500$ Myr. Each fit uses an age prior spanning its panel window. The target masses lie within the emulator’s training range of $1.4,2.5,\mathrm{M}_\odot$. The inference mass prior covers the same range. The metallicity prior is the Soderblom-like form used for the real stars, centred on each target's truth, and rotation is uniform on $[0, 200]\,\mathrm{km\,s^{-1}}$. Whole held-out tracks are limited, so a track may supply models in more than one panel. 400 targets are drawn from 112 distinct tracks. 

Each target was fitted seven times, three emu draws, one hare fit and three noisy-hare draws, for 2800 fits in total. The synthetic mode table for each target is drawn to resemble the real \textit{TESS} detections rather than the full model output: between 20 and 30 modes (median 25) selected from $\ell = 0$--$3$ p\:modes with $n \leq 9$, the $n_{\rm pg} = 0$ f\:modes and the $n_{\rm pg} = -1$ g\:modes at $\ell = 2$ and $3$, with degrees allocated in the proportions observed in typical \dsct stars. Pulsation mode observational uncertainties were drawn from a log-normal fitted to the quoted frequency errors of the five Pleiades targets, with median $5.2\times10^{-5}$\,d$^{-1}$ and a 16--84 per cent range of $8.5\times10^{-6}$ to $3.0\times10^{-4}$\,d$^{-1}$.

For every fit we computed the z-score (per-parameter standardised residual) $z = (\langle\theta\rangle - \theta_{\rm true})/\sigma_\theta$, where $\langle\theta\rangle$ and $\sigma_\theta$ are the posterior mean and standard deviation, so that $z > 0$ means the posterior over-estimates the parameter. For a correctly calibrated pipeline, the 68 and 95 per cent credible intervals ($1\sigma$ and $2\sigma$) contain the truth in 68 and 95 per cent of fits, respectively. For the diagnostics relating to outliers in age recovery, we selected targets whose median $|z_\tau|$ over their four hare and noisy-hare fits exceeds 3. In this campaign this selects the same targets as requiring at least one fit with $|z_\tau|>3$.

Seven fits required more than the $10^7$ likelihood evaluations initially allowed. These fits belonged to two pre-MS targets in the $1$–$50$\,Myr panel, whose age posteriors were $0.08$–$0.17$\,Myr wide within a $49$\,Myr prior. They were rerun with that ceiling raised to $5\times10^{7}$ and converged at $1.3$--$2.3\times10^{7}$ calls. All 2800 fits converged within their assigned budgets and passed the sampler's convergence test. The effective sample size had a minimum of 7058 and a median of 10821.

\begin{table*}
    \centering
    \caption{Population recovery for the hounds campaign. Each entry gives the empirical coverage of the nominal 68 and 95 per cent credible intervals, averaged first over the repeated fits of each target and then over targets, and written as 68/95. $N$ is the number of completed targets. The final column, $\sigma_z(\tau)$, is the standard deviation of the pooled hare and noisy-hare age standardised residuals; unity is the Gaussian calibration reference. The hares and noisy hares are pooled because their measured calibration is similar.}
    \label{tab:hounds}
    \begin{tabular}{lrccccccccr}
        \hline
         & & \multicolumn{4}{c}{Emus} & \multicolumn{4}{c}{Hares and Noisy hares} & \\
        \cmidrule(lr){3-6}\cmidrule(lr){7-10}
        Age panel [Myr] & $N$ & $M$ & $Z$ & $\tau$ & $v_{\rm eq}$ & $M$ & $Z$ & $\tau$ & $v_{\rm eq}$ & $\sigma_z(\tau)$ \\
        \hline
        $1$--$50$ & 100 & 0.68/0.95 & 0.72/0.95 & 0.70/0.95 & 0.68/0.94 & 1.00/1.00 & 1.00/1.00 & 1.00/1.00 & 1.00/1.00 & 0.28 \\
        $50$--$200$ & 100 & 0.67/0.95 & 0.71/0.96 & 0.69/0.95 & 0.69/0.95 & 0.97/1.00 & 0.83/0.97 & 0.78/0.88 & 0.99/1.00 & 1.33 \\
        $200$--$500$ & 100 & 0.68/0.94 & 0.65/0.96 & 0.70/0.95 & 0.66/0.94 & 0.83/0.95 & 0.56/0.78 & 0.39/0.66 & 0.91/0.99 & 2.93 \\
        $500$--$1500$ & 100 & 0.68/0.94 & 0.72/0.97 & 0.67/0.96 & 0.70/0.95 & 0.82/0.96 & 0.58/0.80 & 0.72/0.89 & 0.76/0.97 & 1.41 \\
        \hline
        All & 400 & 0.68/0.95 & 0.70/0.96 & 0.69/0.95 & 0.68/0.95 & 0.90/0.98 & 0.74/0.89 & 0.72/0.86 & 0.91/0.99 & 1.76 \\
        \hline
    \end{tabular}
\end{table*}

\begin{figure*}
    \centering
    \includegraphics[width=\textwidth]{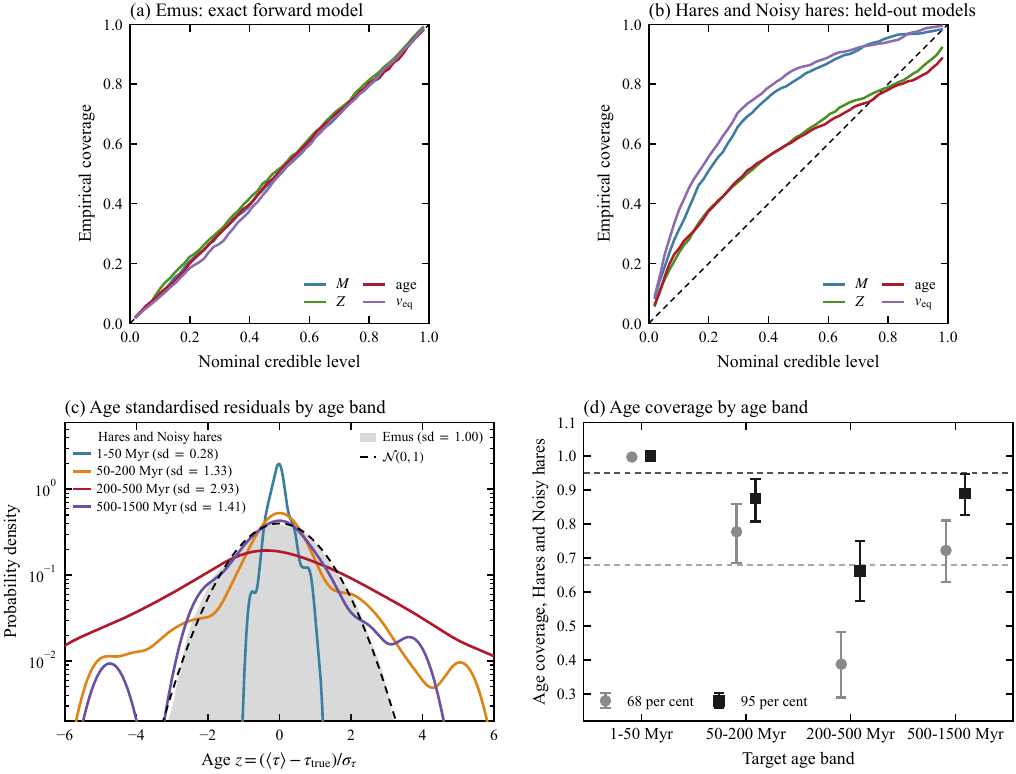}
    \caption{Calibration of the inference pipeline over the emulator's full age domain, from 2800 fits of 400 held-out targets. \emph{(a)} Empirical coverage against nominal credible level for the emus. 
    \emph{(b)} The same for the hares and noisy hares. 
    \emph{(c)} Distribution of the age standardised residual $z=(\langle \tau\rangle-\tau_{\rm true})/\sigma_\tau$ on a logarithmic density scale, for the hares and noisy hares pooled and split by age band, with the emus of all four panels behind them in grey and a unit-normal Gaussian calibration reference (dashed).
    \emph{(d)} Age coverage of the hares and noisy hares in each of the four age panels, with the nominal values as dashed lines. 
    }
    \label{fig:hounds_calibration}
\end{figure*}

\subsubsection{Calibration across the emulator's domain}\label{sec:hh_global}

\begin{figure*}
    \centering
    \includegraphics[width=0.72\textwidth]{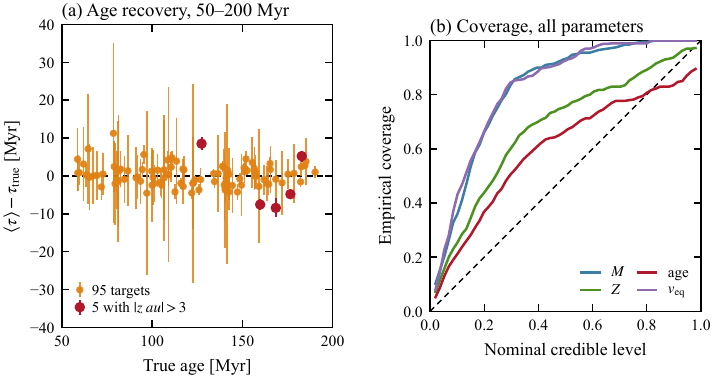}
    \caption{Recovery in the $50$--$200$\,Myr panel, the age window relevant to the Pleiades, from the hares and noisy hares. \emph{(a)} Age residual against true age for the 100 targets, after averaging each target's four posterior means. Error bars show the median posterior standard deviation over those four fits. The five age-outlier targets are highlighted in red: their absolute errors all exceed the panel's 90th percentile, and their posterior standard deviations are about three times smaller than those of the other 95 targets (medians $1.5$ versus $4.8$\,Myr). \emph{(b)} Empirical coverage against nominal credible level for all four parameters, with the dashed diagonal marking nominal coverage. Mass and $v_{\rm eq}$ are conservative, metallicity is close to nominal, and age under-covers above the 80 per cent level.}
    \label{fig:hounds_pleiades}
\end{figure*}

The emus are recovered with near-nominal coverage in all four age panels and for all four parameters. Pooled over the panels, their age coverage is $0.69$ and $0.95$ at the two nominal levels, the standardised residuals have unit standard deviation ($1.00$), and 2 of 1200 fits exceed $|z| = 3$ (Table~\ref{tab:hounds}, Fig.~\ref{fig:hounds_calibration}a and c). Because the emus are generated and fitted with the same emulator and error model, their recovery checks the implementation of the likelihood, covariance treatment and the sampler. 

For the hares and noisy hares, mass and rotation are recovered conservatively. Their coverage curves lie above the diagonal throughout (Fig.~\ref{fig:hounds_calibration}b), and their standardised residuals have standard deviations of $0.64$ and $0.57$. Metallicity and age have residual standard deviations of $1.42$ and $1.76$, and their coverage curves cross the diagonal near the 80 per cent level. They are conservative at low credible levels and under-cover at high ones, consistent with the narrow core and heavy tails seen in Fig.~\ref{fig:hounds_calibration}(c). Excluding fits beyond $|z| = 3$, which are 7.4 per cent of the hare and noisy-hare fits, returns the age residual standard deviation to $0.94$. The hares and noisy hares have the same age residual standard deviation to two decimal places and are therefore pooled in the summaries.

Held-out-grid age recovery depends strongly on evolutionary phase. The target-averaged empirical age coverage is lowest in the $200$--$500$\,Myr panel, at $0.39$ and $0.66$ for nominal 68 and 95 per cent intervals, and remains below nominal in the $500$--$1500$\,Myr panel, at $0.72$ and $0.89$. The failures are concentrated more tightly at $0.77\leq X_{\rm c}/X_0\leq0.91$, owing to the onset of avoided crossings (Appendix~\ref{app:recovery_failures}; Fig.~\ref{fig:hounds_failure_locus}). This $X_{\rm c}/X_0$ interval contains 19 per cent of all targets and 80 per cent of those with $|z_\tau|>3$. The same targets show coupled age and metallicity recovery errors along the population age--metallicity anti-correlation, rather than a separate age-only failure mode. The five Pleiades stars lie at $X_{\rm c}/X_0=0.963$--$0.974$, outside this interval. Appendix~\ref{app:recovery_failures} gives the target-level residual distributions, physical-unit errors, outlier diagnostics and evolutionary-state analysis.

\subsubsection{Calibration in the Pleiades age window}\label{sec:hh_pleiades}

The $50$--$200$\,Myr panel contains the ages reported in Section~\ref{sec:pleiades}, and we treat it separately (Fig.~\ref{fig:hounds_pleiades}). 

Mass, rotation and metallicity have coverage above the nominal levels (Table~\ref{tab:hounds}). Age coverage is $0.777$ $[0.689, 0.861]$ and $0.875$ $[0.808, 0.935]$ at the nominal 68 and 95 per cent levels, respectively. The $1\sigma$ interval is conservative while the $2\sigma$ interval under-covers. The age residuals have a standard deviation of $1.33$ and a mean of $+0.02$, so the pooled residual distribution is centred close to zero despite its broad tails.

Five of the 100 targets carry the whole $|z_\tau|>3$ tail, contributing 20 of the 400 hare and noisy-hare fits. After averaging each target's four posterior means, the median absolute age error is $1.9$\,Myr and the 90th percentile is $4.5$\,Myr. The five outliers have absolute age errors of $4.8$--$8.5$\,Myr ($2.7$--$6.7$ per cent) and narrower posteriors. Their median posterior standard deviation is $1.5$\,Myr, compared with $4.8$\,Myr for the other 95 targets. At the two nominal levels, age coverage is $0.869/0.923$ for the 84 targets outside the $X_{\rm c}/X_0$ window identified above and $0.297/0.625$ for the 16 inside.

\section{Application to $\delta$\,Scuti stars in the Pleiades}
\label{sec:pleiades}

We applied the emulator and inference framework to \dsct stars in the Pleiades observed by \textit{TESS}. From the slow- to moderately-rotating Pleiades \dsct stars presented by \citet{bedding_pleiades_2023}, we selected five whose mode identifications we judged sufficiently reliable for this analysis: V624\,Tau, V647\,Tau, V1187\,Tau, HD\,23157 and HD\,23325.

\begin{table*}
    \centering
    \caption{Observational properties of the five targets. Published spectroscopic effective temperatures and projected rotational velocities are listed alongside the values adopted in this work. The adopted $T_{\rm eff}$ is the only classical observable entering the likelihood, with the quoted uncertainty. The adopted $v\sin i$ is never fitted, but informs the mode identification (Section~\ref{sec:modeid}) and yields the inclination \textit{a posteriori} (Section~\ref{sec:posteriors}).}
    \label{tab:obs_properties}
    \begin{tabular}{lllll}
        \toprule
        Star & $T_{\rm eff}$ published (K) & $T_{\rm eff}$ adopted (K) & $v\sin i$ published (km\,s$^{-1}$) & $v\sin i$ adopted (km\,s$^{-1}$) \\
        \midrule
        V624\,Tau  & 7600--8070$^{b}$ & $7960 \pm 250$ & 32--70$^{b}$; $32.5 \pm 2$$^{a}$; $42 \pm 5$$^{b}$ & $32.5 \pm 2$$^{a}$ \\
        V647\,Tau  & 8055$^{a}$; 8090--8100$^{b}$ & $8055 \pm 250$ & $\leq12$$^{b}$; $10 \pm 3$$^{b}$; $18.9 \pm 1$$^{a}$; $19 \pm 1$$^{c}$ & $18.9 \pm 1$$^{a}$ \\
        V1187\,Tau & 8031$^{d}$ & $8031 \pm 250$ & $37 \pm 1$$^{c,d}$ & $37 \pm 1$$^{c,d}$ \\
        HD\,23157  & 7514$^{a}$ & $7514 \pm 250$ & 56$^{a}$; $58 \pm 1$$^{c}$ & $58 \pm 1$$^{c}$ \\
        HD\,23325  & 7638$^{a}$ & $7638 \pm 250$ & 80$^{a}$; $85 \pm 1$$^{c}$ & $85 \pm 1$$^{c}$ \\
        \bottomrule
    \end{tabular}
    \\[2pt]
    \raggedright
    \footnotesize
    $^{a}$\citet{gebran_chemical_2008}; $^{b}$\citet{murphy_five_2022}, and references therein; $^{c}$\citet{bedding_pleiades_2023}; $^{d}$\citet{prugniel_atmospheric_2011}.
\end{table*}

\subsection{Targets and observations}\label{sec:targets}

Table~\ref{tab:obs_properties} lists the published and adopted effective temperatures and projected rotational velocities. For all five targets, we extracted the \textit{TESS} light curves using \textsc{lightkurve} and measured frequencies by iterative prewhitening\footnote{\href{https://github.com/gautam-404/prewhitener}{github.com/gautam-404/prewhitener}}. Table~\ref{tab:obs_tess} lists the sectors observed and the resulting light-curve baselines. Four of the targets were observed in Sectors 42--44 and again in Sectors 70 and 71, giving a baseline of 812\,d. V647\,Tau has two-minute data only in Sectors 42--44, so its baseline is 77\,d. Its median frequency uncertainty is more than an order of magnitude larger than that of any other target. 

\begin{table}
    \centering
    \footnotesize
    \setlength{\tabcolsep}{5pt}
    \caption{\textit{TESS} observations of the five targets. All five light curves are two-minute cadence SPOC photometry. $N_{\rm ext}$ is the number of significant frequencies recovered by prewhitening and $N_{\rm fit}$ the number of identified modes entering the likelihood (Appendix~\ref{app:modeids}). No two-minute observations of V647\,Tau are available beyond Sector~44.}
    \label{tab:obs_tess}
    \begin{tabular}{llcrrr}
        \toprule
        Star & TIC & Sectors & Baseline & $N_{\rm ext}$ & $N_{\rm fit}$ \\
             &     &         & (d)      &               &              \\
        \midrule
        V624\,Tau  & 405483817 & 42--44, 70, 71 & 811.7 & 117 & 26 \\
        V647\,Tau  & 405484188 & 42--44         &  76.8 &  60 & 20 \\
        V1187\,Tau & 405483707 & 42--44, 70, 71 & 811.7 & 101 & 22 \\
        HD\,23157  &  67768222 & 42--44, 70, 71 & 811.7 &  72 & 19 \\
        HD\,23325  & 385509282 & 42--44, 70, 71 & 811.7 & 122 & 16 \\
        \bottomrule
    \end{tabular}
\end{table}

V624\,Tau was observed at two-minute cadence across multiple sectors. We recovered 117 significant frequencies and fitted 26 identified modes in the 18--63\,d$^{-1}$ range (Table~\ref{tab:modeids_v624}): 19 p\:modes spanning $\ell=0$--$3$, six f\:modes (one $\ell=2$ and five $\ell=3$), and one $\ell=3$ g\:mode. Several belong to rotationally split multiplets. Its large separation is $\Delta\nu=7.00$\,d$^{-1}$. The omitted $n=4$ radial position is discussed in Section~\ref{sec:posteriors}.

V647\,Tau has 20 fitted modes in the 18--59\,d$^{-1}$ range (Table~\ref{tab:modeids_v647}): 15 p\:modes (five radial, eight dipole and two $\ell=2$), three $\ell=3$ f\:modes, and two $\ell=2$ g\:modes. Its large separation is $\Delta\nu=6.88$\,d$^{-1}$. The radial ridge skips $n=3$ and $4$ orders. The strongest peak lies near the expected $n=4$ position but is excluded from the fit (Section~\ref{sec:posteriors}).

V1187\,Tau (HD\,23194) has 22 fitted modes in the 20--71\,d$^{-1}$ range (Table~\ref{tab:modeids_v1187}): 16 p\:modes (four radial, ten dipole and two $\ell=3$), three $\ell=2$ f\:modes, and three $\ell=3$ g\:modes. Its large separation is $\Delta\nu=7.02$\,d$^{-1}$.

HD\,23157 (TIC\,67768222) has 19 fitted modes spanning 17.0--55.2\,d$^{-1}$ (Table~\ref{tab:modeids_hd23157}): 13 p\:modes (two radial, two dipole, four $\ell=2$ and five $\ell=3$), and six low-order $n_{\rm pg}=-1$ g\:modes. The latter are the even-parity components of the $\ell=2$ and $\ell=3$ multiplets around $3\Delta\nu$, with $\Delta\nu=6.85$\,d$^{-1}$.

HD\,23325 is the only chemically peculiar target, a metallic-line (Am) star classified kA6hA9mF1\,(IV). Its 16 fitted modes span 15.6--65.8\,d$^{-1}$ (Table~\ref{tab:modeids_hd23325}): 13 p\:modes (six radial and seven dipole) and three low-order $\ell=2$ g\:modes below the fundamental radial mode. Its large separation is $\Delta\nu=6.90$\,d$^{-1}$.

\subsection{Mode identification}\label{sec:modeid}

Before any inference is run, each star's pulsation spectrum must be reduced to a list of securely labelled $(n, \ell, m)$ modes (Section~\ref{sec:inference_likelihood}). We built this list by hand from the \'echelle diagram, following a consistent procedure for all five stars. The resulting identifications are tabulated in Appendix~\ref{app:modeids}.

We began by searching for the large separation $\Delta\nu$ for which the radial ($\ell=0$) and dipole ($\ell=1$) modes with radial orders $n>3$ form approximately vertical ridges in the \'echelle diagram. This regular high-frequency pattern is the pseudo-asymptotic p-mode structure resolved in \'echelle diagrams of young \dsct stars with space photometry \citep{bedding_very_2020, murphy_precise_2021, murphy_five_2022}. At the modest rotation rates of our targets the ridges are distorted from the non-rotating asymptotic form, so the pattern is sought as approximately, rather than exactly, vertical \citep{reese_modelling_2008, ouazzani_pulsations_2015}.

With a trial $\Delta\nu$, we first anchored the radial ridge using the higher-order radial modes ($n \gtrsim 5$). These form a near-vertical ridge at the correct $\Delta\nu$, while the lowest radial orders curve to the right from the vertical ridge and are added only once the ridge is fixed. The fundamental radial mode ($n=1$) is expected near $3\Delta\nu$ \citep{gautam_modelling_2025}, providing a reference for the radial-order numbering.

We next identified the non-radial modes, starting with the central ($m=0$) components where they are detectable. The $\ell=1$ central ridge lies to the left of the radial ridge, while the $\ell=2$ central modes lie between them.

The detectability of these components depends on the stellar inclination through their disc-integrated visibility $V_{\ell m}(i)$ (Fig.~\ref{fig:mode_visibility}; \citealp{gizon_determining_2003, aerts_asteroseismology_2010, reese_2d_2022}). In this geometric model, the $m=0$ visibility vanishes at $i=90^\circ$ (equator-on) for $\ell=1$, near $54.7^\circ$ for $\ell=2$, and near $39^\circ$ and $90^\circ$ for $\ell=3$. A central component may therefore be absent because of geometric cancellation, but this alone does not identify a particular inclination. We used the visibility curves and spectroscopic $v\sin i$ to guide the mode assignments, but these constraints do not uniquely determine the labels.

Finally, we identify the rotationally split multiplets. The frequency spacing of a multiplet constrains the rotation rate, so we require any candidate splitting to be consistent with the spectroscopic $v\sin i$. Because $\sin i \leq 1$, the projected velocity $v\sin i$ is a lower bound on the equatorial velocity $v_{\rm eq}$, and the observed splitting should therefore match or exceed the spacing predicted from $v\sin i$ at the model radius. Splittings narrower than this bound are not considered during mode identification. The inferred equatorial velocity can then be combined with $v\sin i$ to obtain an inclination \textit{a posteriori}.

The $(n,\ell,m)$ labels are fixed before inference, and neither inclination nor mode identity is sampled in the likelihood. We assessed a limited set of mode tables separately through their per-mode residuals and fitted \'echelle diagrams.

We assign $(n,\ell,m)$ labels according to the dominant non-rotating component of each observed mode. Centrifugal distortion can couple modes of similar frequency with the same $m$ and $\Delta\ell=0,\pm2$ \citep{dziembowski_effects_1992, soufi_effects_1998, suarez_role_2007, guo_oscillation_2024}. For radial modes, the relevant partner is the $\ell=2,m=0$ component. Near-degeneracy may become important when their uncoupled frequencies satisfy $|\nu_{\ell=0}-\nu_{\ell=2,m=0}|\lesssim\nu_{\rm rot}$, where $\nu_{\rm rot}$ is the rotation frequency. The amount of mixing also depends on the coupling strength, so frequency proximity alone does not establish a mixed identity. Peaks whose dominant degree cannot be identified are left unlabelled and excluded from the fit.

A further consequence of nonlinear mode interaction is that not every peak in the amplitude spectrum is an independent eigenmode. Pairs of strong modes generate combination frequencies (sums, differences and harmonics of the dominant pulsations) that appear as sharp peaks but carry no independent $(n, \ell, m)$ label \citep{mourabit_resonant_2023}. Our iterative prewhitening flags integer harmonics ($\nu \approx k\,\nu_0$) automatically, and any flagged harmonic, together with any other peak we cannot securely identify as an eigenmode, is excluded from the fit rather than assigned a spurious label.

All five stars show strong peaks near the fundamental radial-mode frequency, around $3\Delta\nu$. Secure identifications in this region include low-order g\:modes and f\:modes, with several peaks below the lowest identified radial mode.

In V624\,Tau, an $\ell = 3$ g\:mode at $18.81$\,d$^{-1}$ and an $\ell = 2$ f\:mode at $20.98$\,d$^{-1}$ lie below the lowest identified radial mode ($n = 2$, at $27.46$\,d$^{-1}$). V647\,Tau has two $\ell = 2$ g\:modes at $17.89$ and $18.69$\,d$^{-1}$, below its radial fundamental at $20.85$\,d$^{-1}$. V1187\,Tau has three $\ell = 3$ g\:modes spanning $19.81$--$20.81$\,d$^{-1}$ and three identified components of an $\ell = 2$ f\:mode multiplet spanning $22.38$--$23.48$\,d$^{-1}$. The richest set occurs in HD\,23157, where six components of the $n_{\rm pg} = -1$, $\ell = 2$ and $\ell = 3$ multiplets span $17.0$--$23.7$\,d$^{-1}$, straddling the radial fundamental at $21.00$\,d$^{-1}$. HD\,23325 has three $\ell = 2$ g\:modes spanning $15.6$--$18.3$\,d$^{-1}$.

The identified g\:modes have low radial orders ($|n_{\rm pg}| \leq 2$), unlike the high-order g\:modes characteristic of $\gamma$~Doradus pulsators; their presence alone therefore does not establish hybrid pulsation. Peaks without secure identifications are excluded from the fits, regardless of amplitude. In V647\,Tau, for example, the second-strongest peak in the entire amplitude spectrum remains unidentified at $20.44$\,d$^{-1}$, just $0.41$\,d$^{-1}$ below the radial fundamental.

Several diagnostics can flag a possible mismatch between the adopted mode identifications and the model grid. These include anomalous per-mode frequency residuals, a high reduced frequency $\chi^2_{\nu,\rm freq}$, a peak's position in the \'echelle diagram, a $>3 \sigma$ disagreement between posterior-predicted and spectroscopic observables, or a posterior pressed against a prior or grid boundary. None of these diagnostics uniquely identifies a wrong mode label. Similar patterns can arise from inadequate input physics, an inappropriate prior, an emulator limitation, or another form of model--data mismatch. We treat these diagnostics as a reason to investigate the adopted mode identifications.

In HD\,23157 these diagnostics led us to exclude three modes. We ranked its 22 identified modes by the change in frequency $\chi^2$ when each mode is removed individually. We calculated these changes analytically from the full covariance matrix at the posterior mean of the 22-mode fit, without refitting. We excluded the three highest-ranked modes, listed as X1--X3 in Table~\ref{tab:modeids_hd23157}. Refitting the remaining 19 modes lowered $\chi^2_{\nu,\rm freq}$ from $9.4$ to $5.4$ and shifted the median age by $2.3$\,Myr, well below its $1\sigma$ uncertainty. This selection is based on fit quality rather than evidence of incorrect identifications.

\subsection{Posterior summaries for individual stars}\label{sec:posteriors}

\begin{table*}
    \centering
    \caption{Posterior summaries for the five Pleiades targets. The four fitted parameters and the two derived observables are reported as posterior medians with $1\sigma$ (16th--84th-percentile) intervals. $T_{\rm eff}$ and $\log_{10}(L/\mathrm{L}_\odot)$ are not sampled; instead, they are obtained by evaluating the emulator at every posterior sample. The effective temperature is additionally constrained by the classical term of the likelihood, through the adopted values of Table~\ref{tab:obs_properties}, whereas the luminosity is predicted by the fitted models alone. The reduced frequency $\chi^2_{\nu,\rm freq}$ is evaluated at the posterior median against the full covariance $\mathrm{C}_\nu$. The inclination $i$ is derived after the fit from the asteroseismic $v_{\rm eq}$ and the spectroscopic $v\sin i$. $\Omega/\Omega_{\rm crit}$ is the posterior median of \mesa's surface-averaged ratio, evaluated on the same samples as $T_{\rm eff}$ and $\log_{10}(L/\mathrm{L}_\odot)$.}
    \label{tab:pleiades_posteriors}
    \renewcommand{\arraystretch}{1.3}
    \setlength{\tabcolsep}{5.5pt}
    \begin{tabular}{lccccccccc}
        \toprule
        Star & $M$ ($\mathrm{M}_\odot$) & $Z$ & Age (Myr) & $v_{\rm eq}$ (km\,s$^{-1}$) & $\Omega/\Omega_{\rm crit}$ & $T_{\rm eff}$ (K) & $\log_{10}(L/\mathrm{L}_\odot)$ & $\chi^2_{\nu,\rm freq}$ & $i$ (deg) \\
        \midrule
        \addlinespace[2pt]
        V624\,Tau  & $1.696^{+0.005}_{-0.005}$ & $0.0179^{+0.0002}_{-0.0002}$ & $116.9^{+7.7}_{-8.2}$ & $48.1^{+0.2}_{-0.2}$ & 0.105 & $7988^{+29}_{-30}$ & $0.934^{+0.007}_{-0.007}$ & 9.1 & $42.5 \pm 3.2$ \\
        \addlinespace[5pt]
        V647\,Tau  & $1.721^{+0.006}_{-0.006}$ & $0.0192^{+0.0002}_{-0.0002}$ & $113.7^{+9.3}_{-9.6}$ & $40.3^{+0.2}_{-0.2}$ & 0.088 & $7972^{+36}_{-36}$ & $0.946^{+0.009}_{-0.009}$ & 14.0 & $28.0 \pm 1.6$ \\
        \addlinespace[5pt]
        V1187\,Tau & $1.665^{+0.011}_{-0.010}$ & $0.0177^{+0.0003}_{-0.0004}$ & $112.6^{+17.5}_{-17.4}$ & $40.4^{+0.3}_{-0.3}$ & 0.089 & $7880^{+66}_{-64}$ & $0.901^{+0.016}_{-0.016}$ & 11.9 & $66.3 \pm 3.6$ \\
        \addlinespace[5pt]
        HD\,23157  & $1.514^{+0.007}_{-0.006}$ & $0.0189^{+0.0001}_{-0.0002}$ & $114.6^{+13.8}_{-12.2}$ & $76.7^{+0.2}_{-0.2}$ & 0.175 & $7115^{+36}_{-28}$ & $0.708^{+0.010}_{-0.008}$ & 5.4 & $49.1 \pm 1.1$ \\
        \addlinespace[5pt]
        HD\,23325  & $1.639^{+0.008}_{-0.009}$ & $0.0181^{+0.0002}_{-0.0002}$ & $107.0^{+9.3}_{-10.6}$ & $89.1^{+0.4}_{-0.5}$ & 0.198 & $7701^{+47}_{-48}$ & $0.865^{+0.012}_{-0.012}$ & 10.5 & $72.6 \pm 2.3$ \\
        \addlinespace[2pt]
        \bottomrule
    \end{tabular}
\end{table*}

\begin{figure}
    \centering
    \begin{minipage}{\linewidth}
        \centering
        \includegraphics[width=0.93\linewidth]{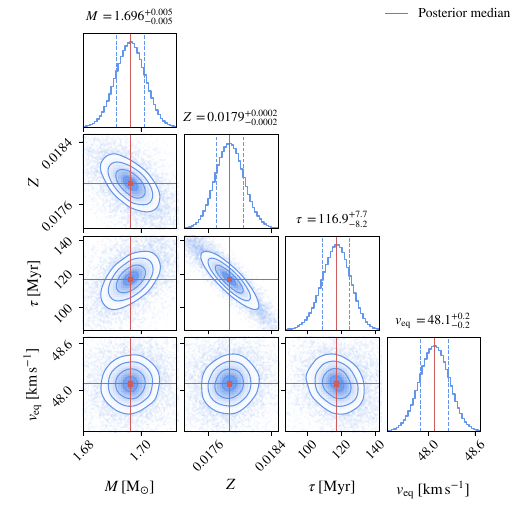}\\[1em]
        \includegraphics[width=0.93\linewidth]{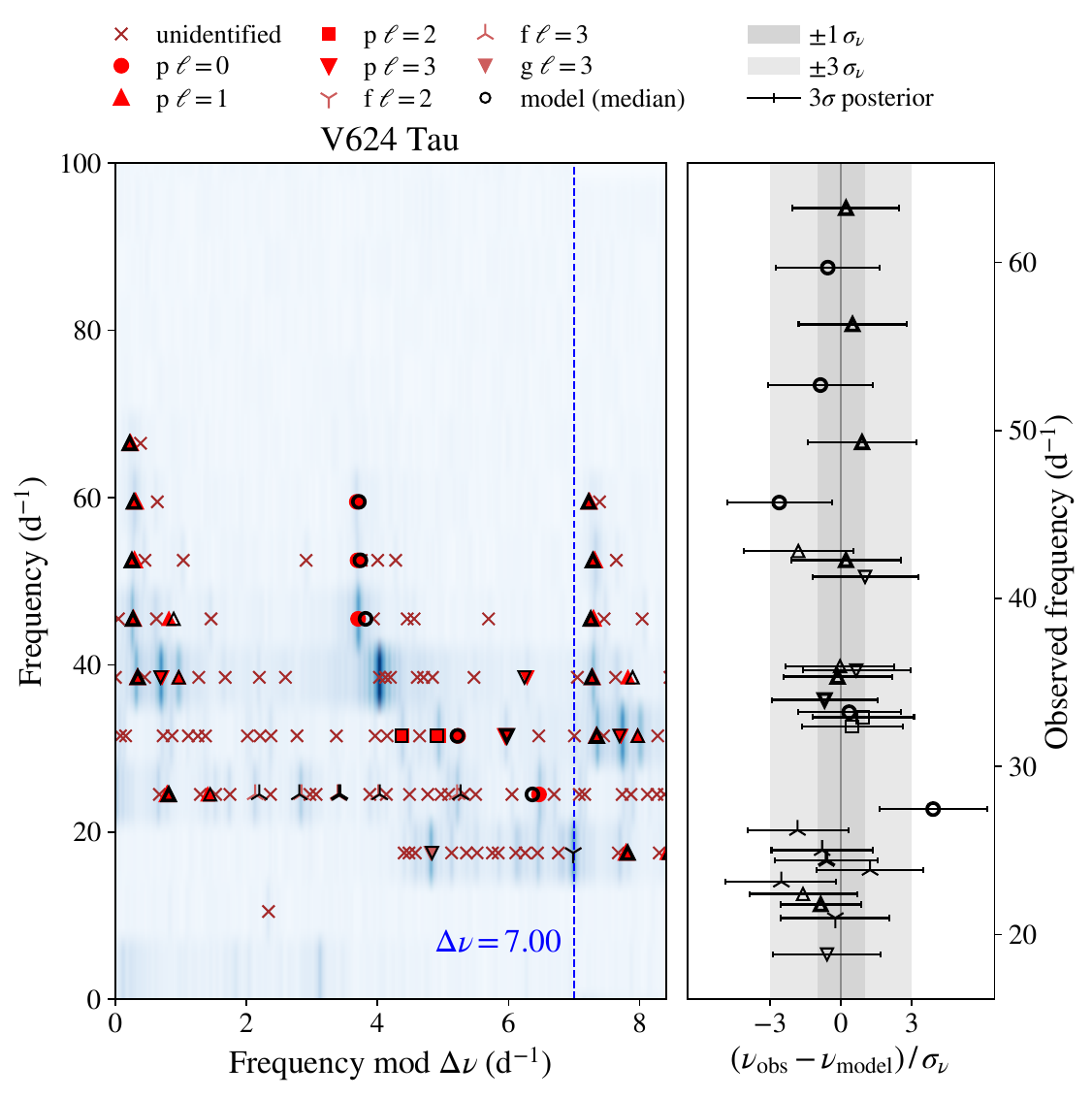}
    \end{minipage}
    \caption{Inference results for V624\,Tau. \textbf{Top:} Joint posterior for $M$, $Z$, $\tau$, and $v_{\rm eq}$. Red lines mark the posterior medians, dashed lines the 16th and 84th percentiles, and contours enclose the 0.5, 1, 1.5 and 2$\sigma$ credible regions. \textbf{Bottom:} \'Echelle diagram folded on $\Delta\nu=7.00$\,d$^{-1}$, with normalised frequency residuals. The red symbols mark identified observed peaks, and red crosses mark unidentified peaks. Open black symbols give the median of each fitted mode's frequency over the posterior, with mode families and degrees identified in the legend. The panel to the right shows the residual $\nu_{\rm obs}-\nu_{\rm model}$ of each fitted mode against its observed frequency, in units of $\sigma_\nu=\sqrt{\mathrm{diag}(\mathrm{C}_\nu)}$, the combined observational and emulator error on that mode's frequency (Eq.~\eqref{eq:cov_terms}). Bars here span the central 3$\sigma$ posterior interval on the model frequency, and the nested shading marks $\pm1$ and $\pm3\,\sigma_\nu$. The vertical axes of the \'echelle and residual panels show different quantities: the \'echelle places a mode at its order centre, while the residual panel places it at its own observed frequency. Identified mode frequency labels match Appendix~\ref{app:modeids}.}
    \label{fig:v624tau_results}
\end{figure}

We ran the inference framework of Section~\ref{sec:inference} on each star individually. All parameter uncertainties quoted in this section are \emph{formal}, statistical credible intervals, conditional on the adopted $(n,\ell,m)$ mode identification (Section~\ref{sec:inference_likelihood}). Uncertainty arising from the adopted input physics is not included here and is examined in Section~\ref{sec:input_physics_sensitivity}.

All five stars are members of the Pleiades and are expected to have similar initial compositions, so we adopted the same metallicity prior for every target, based on the cluster measurement of \citet{soderblom_metallicity_2009}. For HD\,23325, the Am star, its metallic-line surface abundances arise from atomic diffusion in a shallow surface layer and do not trace the bulk composition that the pulsation frequencies respond to, so the cluster mean from solar-type members remains the better bulk-composition anchor than any individual A-star spectrum. Uniform priors are used for the remaining sampling parameters. The effective temperature is the only classical observable entering the likelihood, with the adopted values of Table~\ref{tab:obs_properties}. The equatorial velocity is constrained by the identified rotationally split multiplets rather than by any external $v\sin i$ measurement.

\begin{figure}
    \centering
    \begin{minipage}{\linewidth}
        \centering
        \includegraphics[width=0.93\linewidth]{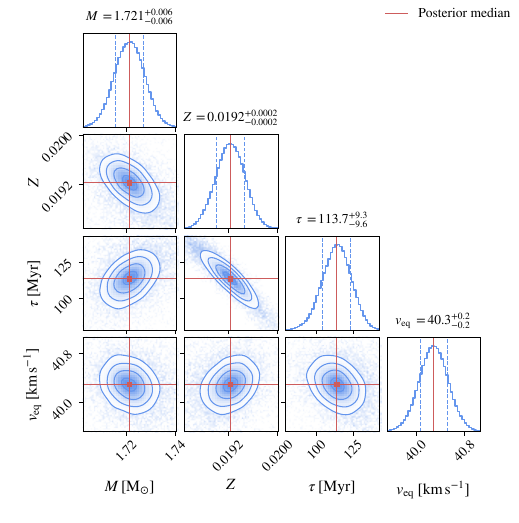}\\[1em]
        \includegraphics[width=0.93\linewidth]{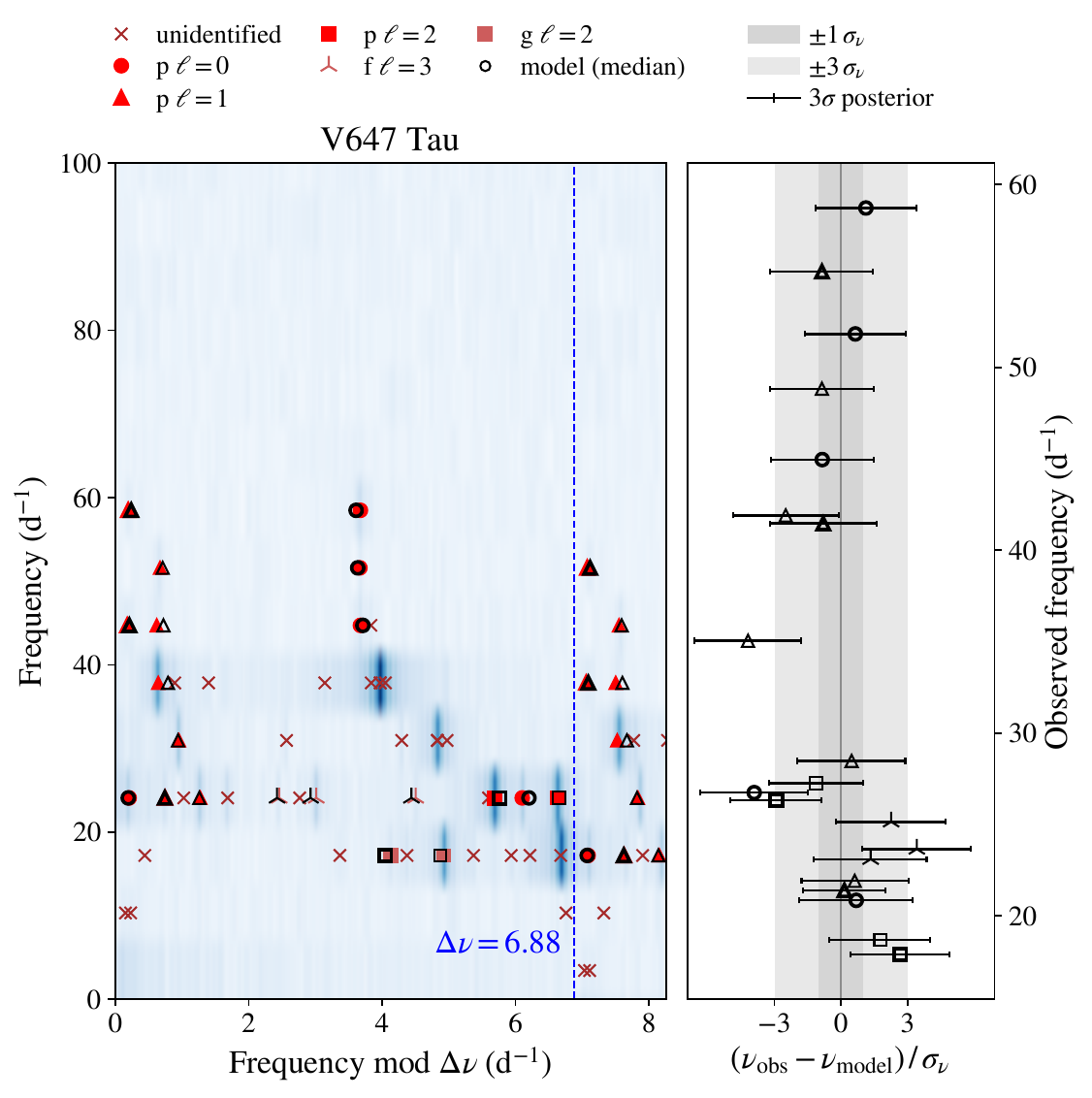}
    \end{minipage}
    \caption{Inference results for V647\,Tau. Posterior, \'echelle and residual conventions follow Fig.~\ref{fig:v624tau_results}.}
    \label{fig:v647tau_results}
\end{figure}

\begin{figure}
    \centering
    \begin{minipage}{\linewidth}
        \centering
        \includegraphics[width=0.93\linewidth]{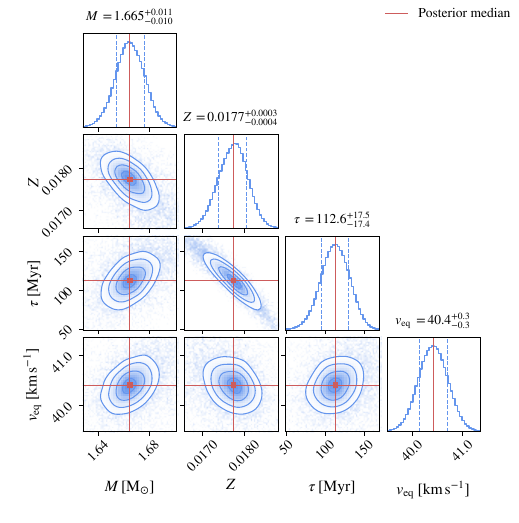}\\[1em]
        \includegraphics[width=0.93\linewidth]{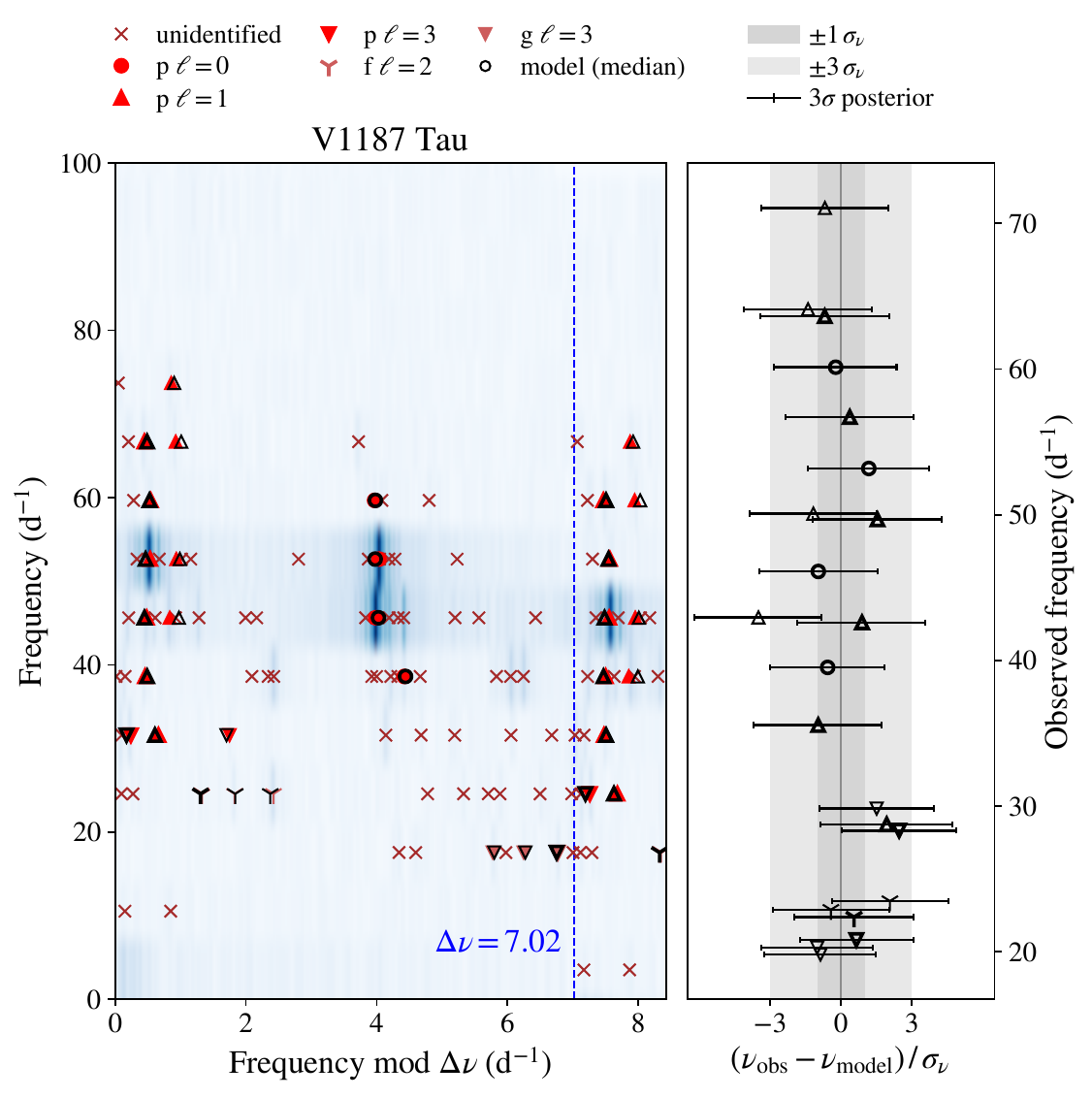}
    \end{minipage}
    \caption{Inference results for V1187\,Tau. Posterior, \'echelle and residual conventions follow Fig.~\ref{fig:v624tau_results}.}
    \label{fig:v1187tau_results}
\end{figure}

\begin{figure}
    \centering
    \begin{minipage}{\linewidth}
        \centering
        \includegraphics[width=0.93\linewidth]{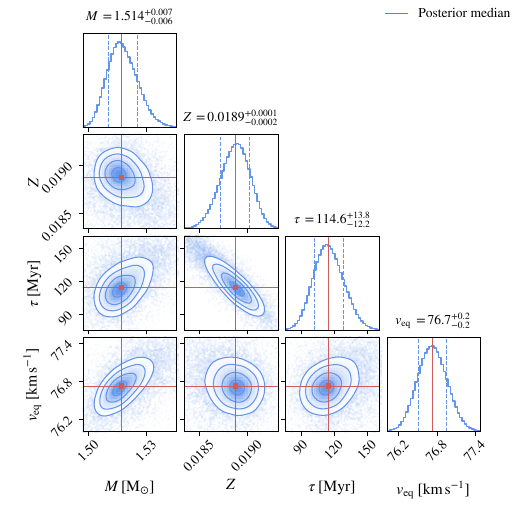}\\[1em]
        \includegraphics[width=0.93\linewidth]{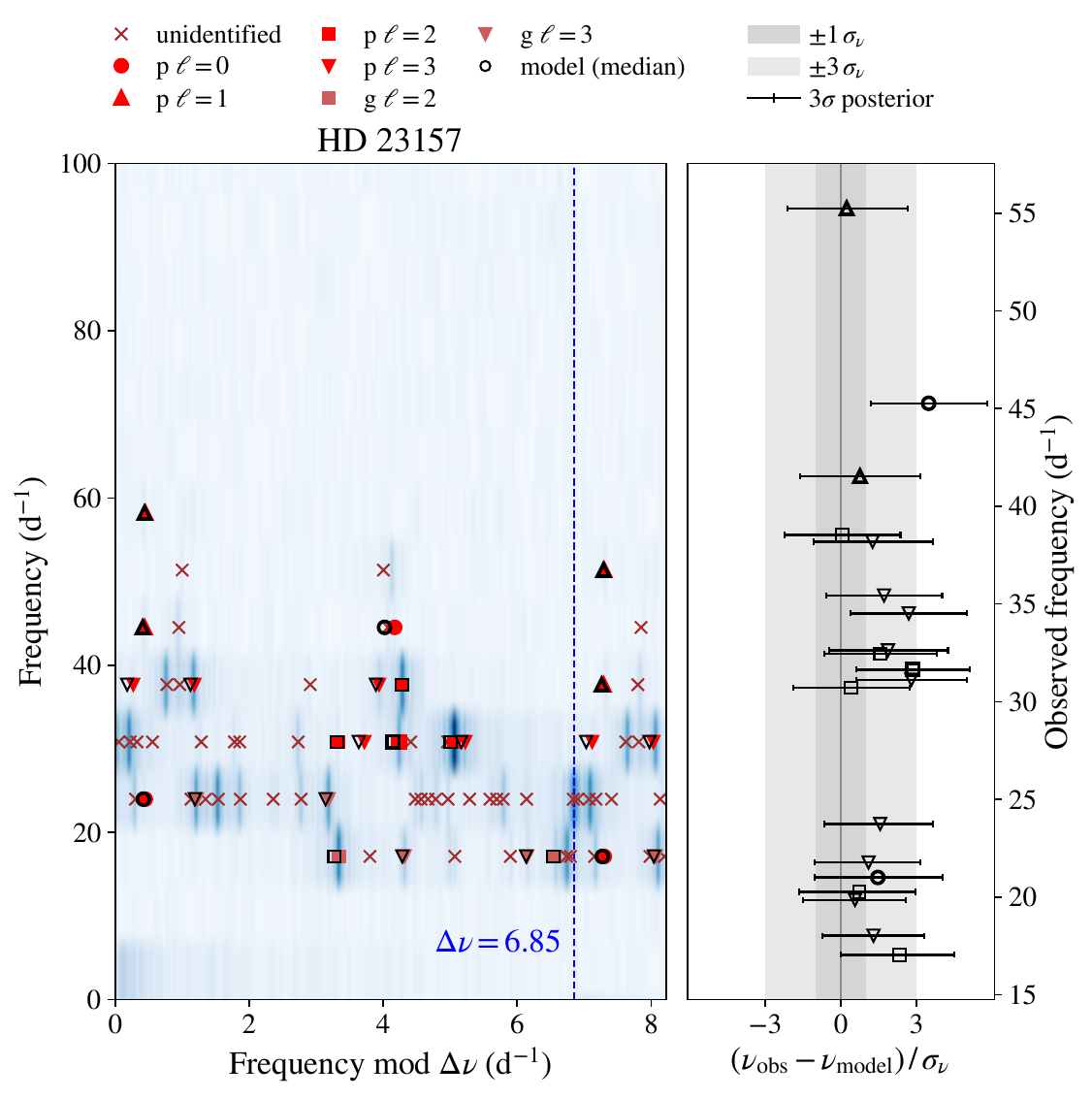}
    \end{minipage}
    \caption{Inference results for HD\,23157. Posterior, \'echelle and residual conventions follow Fig.~\ref{fig:v624tau_results}.}
    \label{fig:hd23157_results}
\end{figure}

\begin{figure}
    \centering
    \begin{minipage}{\linewidth}
        \centering
        \includegraphics[width=0.93\linewidth]{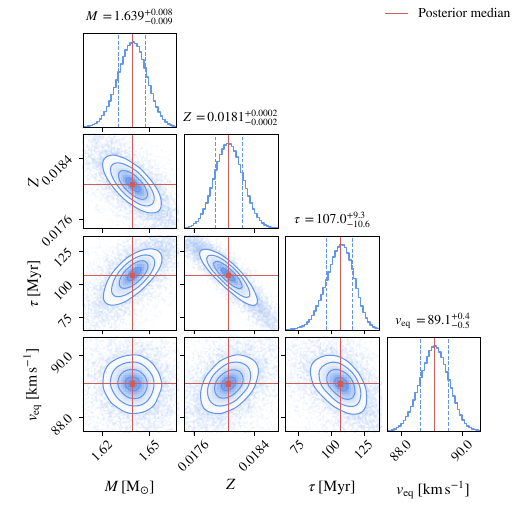}\\[1em]
        \includegraphics[width=0.93\linewidth]{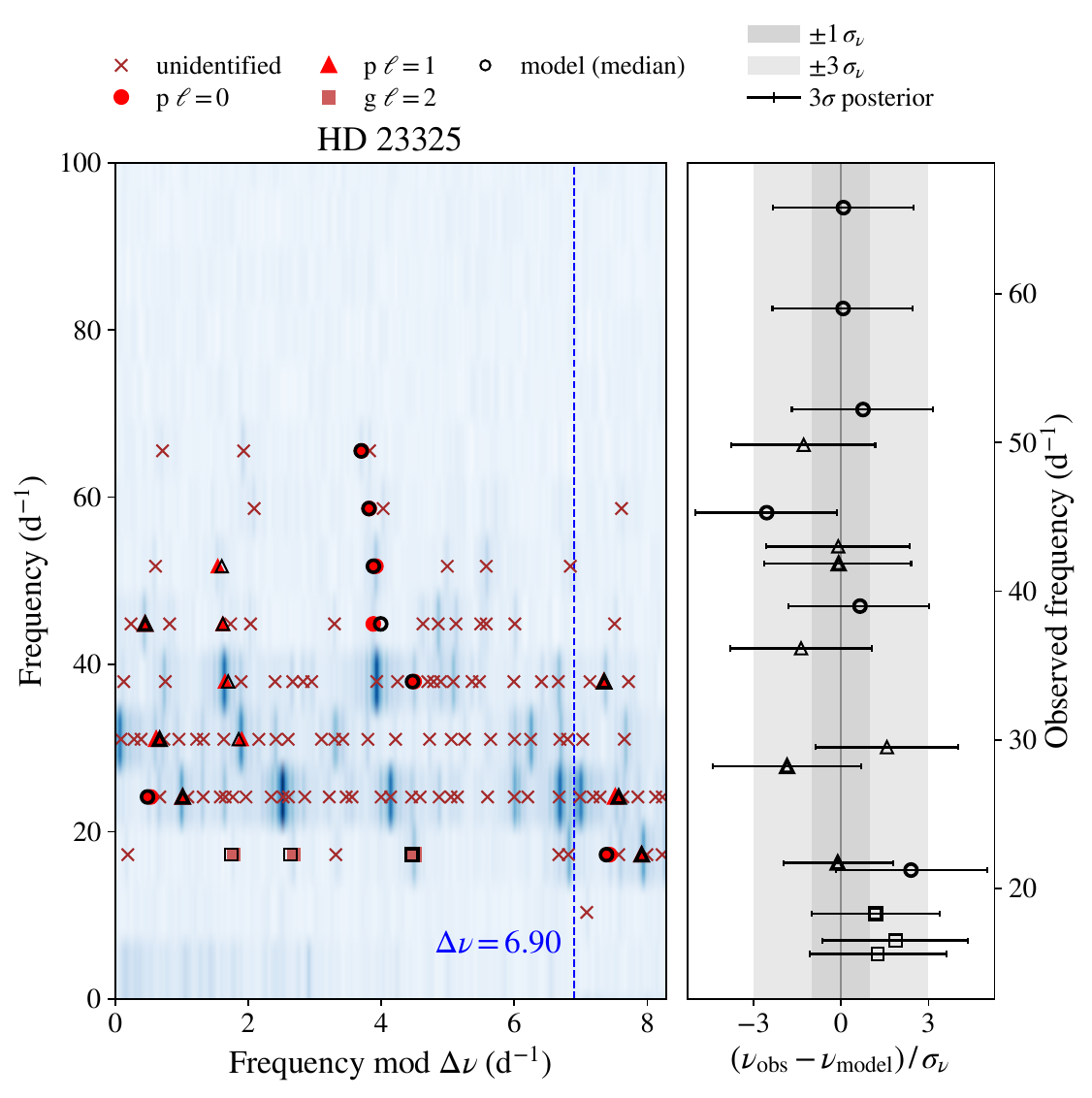}
    \end{minipage}
    \caption{Inference results for HD\,23325. Posterior, \'echelle and residual conventions follow Fig.~\ref{fig:v624tau_results}.}
    \label{fig:hd23325_results}
\end{figure}

Table~\ref{tab:pleiades_posteriors} summarises the inferred parameters. Figures~\ref{fig:v624tau_results}--\ref{fig:hd23325_results} show the corresponding posterior distributions and frequency comparisons. All five stars have posterior median $\Omega/\Omega_{\rm crit} = 0.09$--$0.20$ and $v_{\rm eq} = 40$--$89$\,km\,s$^{-1}$, below both the $0.2$ ratio reached by the only published \storm benchmark and the $100$\,km\,s$^{-1}$ g\:mode bound of Section~\ref{sec:grids}. 

For V624\,Tau, our frequencies and mode labels agree with the \textit{Kepler} K2-mission based radial and dipole identifications of \citet{murphy_five_2022} where they overlap. Literature $v\sin i$ values span $32$--$70$\,km\,s$^{-1}$ (Table~\ref{tab:obs_properties}; \citealp{bedding_pleiades_2023} report no measurement for this star). We adopted the high-resolution determination $32.5\pm2$\,km\,s$^{-1}$ \citep{gebran_chemical_2008}, which gives $i=42.5^\circ\pm3.2^\circ$ through $v_{\rm eq}\sin i=v\sin i$. The inclination uncertainty is dominated by spectroscopy because the asteroseismic $v_{\rm eq}$ is constrained to better than $1$\,per\,cent within the model. 

V624\,Tau's age agrees with the external cluster-age estimates discussed in Section~\ref{sec:cluster_age}, and its metallicity is consistent with the adopted spectroscopic prior.

Our mode identifications for V647\,Tau also agree with the overlapping K2-based radial and dipole identifications of \citet{murphy_five_2022}. The adopted $v\sin i=18.9\pm1$\,km\,s$^{-1}$ \citep{gebran_chemical_2008} implies a low inclination, $i=28.0^\circ\pm1.6^\circ$. The TRES value of $19\pm1$\,km\,s$^{-1}$ \citep{bedding_pleiades_2023} supports this choice over the smaller historical determinations compiled by \citet{murphy_five_2022} (Table~\ref{tab:obs_properties}). 

The V647\,Tau \'echelle diagram (Fig.~\ref{fig:v647tau_results}) shows one ambiguous feature. The strongest peak in the amplitude spectrum lies at $\approx 38.4$\,d$^{-1}$, near the expected $n = 4$ radial position obtained by extrapolating the $n = 5$, 6 and 7 radial orders with $\Delta\nu \approx 6.88$\,d$^{-1}$. A second unidentified peak lies $0.83$\,d$^{-1}$ below it. We tested radial, $\ell = 2$, and $\ell = 3$ interpretations of the stronger peak. The $\ell=2$ and $\ell=3$ interpretations gave slightly better frequency fits (in reduced frequency $\chi^2_{\nu,\rm freq}$) than the radial interpretation, but we could not securely distinguish between them. Both peaks therefore remain unlabelled and are excluded from the adopted fit.

A similar ambiguity occurs at the expected $n = 4$ radial position in V624\,Tau. In both stars the peak is left unlabelled and excluded from the fit.

V1187\,Tau has the broadest age posterior of the five stars, $112.6^{+17.5}_{-17.4}$\,Myr. Its inferred effective temperature agrees with the adopted spectroscopic value. Combining the inferred equatorial velocity with the spectroscopic $v\sin i=37\pm1$\,km\,s$^{-1}$ \citep{bedding_pleiades_2023, prugniel_atmospheric_2011} gives $i=66.3^\circ\pm3.6^\circ$.

HD\,23157 has the lowest inferred mass in the sample, $1.514^{+0.007}_{-0.006}\,\mathrm{M}_\odot$. The adopted spectroscopic $v\sin i=58\pm1$\,km\,s$^{-1}$ \citep{bedding_pleiades_2023} agrees with the $56$\,km\,s$^{-1}$ measurement of \citet{gebran_chemical_2008} and, combined with the inferred equatorial velocity, gives $i=49.1^\circ\pm1.1^\circ$. 

HD\,23325 has the highest inferred equatorial velocity, $89.1^{+0.4}_{-0.5}$\,km\,s$^{-1}$. The \textit{Gaia} DR3 $v\sin i=85\pm1$\,km\,s$^{-1}$ \citep{bedding_pleiades_2023} implies $i=72.6^\circ\pm2.3^\circ$; adopting $80$\,km\,s$^{-1}$ from \citet{gebran_chemical_2008} would instead give $i\approx64^\circ$. Under the flat metallicity prior, its metallicity changes only slightly and its median age decreases by $7.6$\,Myr (Appendix~\ref{app:metallicity_prior}); the Am surface abundances enter neither fit. 

Within the adopted grid, the p\:modes alone constrain a narrow ridge in the age--metallicity plane, and the age constraint tightens further with either a metallicity measurement or an identified g\:mode. All five stars fitted here carry at least one identified low-order g\:mode. Their age and metallicity posteriors remain constrained when the cluster metallicity prior is replaced by a uniform prior, although the individual age intervals broaden by 10--35 per cent (Appendix~\ref{app:metallicity_prior}).

\subsection{The age of the Pleiades}\label{sec:cluster_age}
\setlength{\dbltextfloatsep}{2pt plus 1pt}
\begin{figure*}
    \centering
    \includegraphics[width=\textwidth]{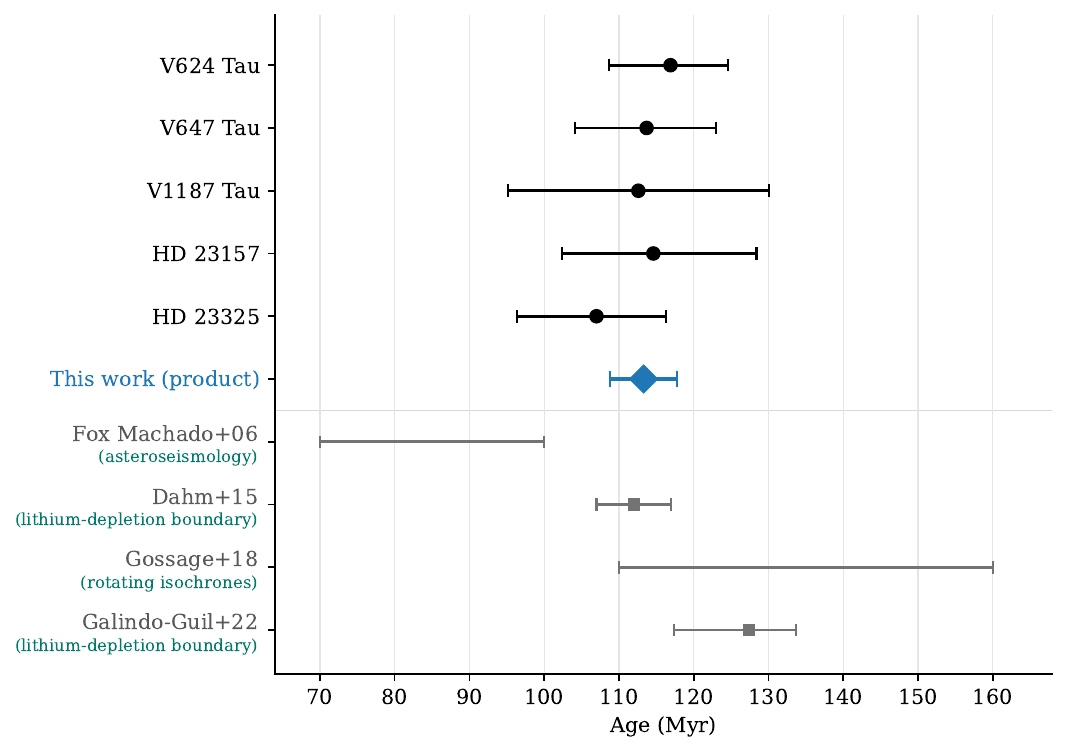}
    \caption{Individual Pleiades age estimates and their combined posterior, compared with literature ages for the cluster. Points show posterior medians and horizontal bars their $1\sigma$  statistical intervals. The blue diamond and error bar show the combined shared-age posterior of Eq.~\eqref{eq:age_product}, $113.3\pm4.5$\,Myr. The grey rows below the rule are literature ages, ordered by publication year: the earlier asteroseismic age from six Pleiades \dsct stars \citep[70--100\,Myr;][]{fox_machado_seismological_2006}, the lithium-depletion boundary \citep[$112 \pm 5$\,Myr;][]{dahm_2015}, rotating-isochrone fits to the colour--magnitude diagram \citep[110--160\,Myr;][]{gossage_pleiades_2018}, and that same lithium-depletion boundary relocated with \textit{Gaia} DR2 parallaxes \citep[$127.4^{+6.3}_{-10}$\,Myr;][]{galindo-guil_lithium_2022}. Rows drawn without a central marker are ranges of acceptable fits rather than $1\sigma$  intervals.}
    \label{fig:cluster_age}
\end{figure*}

Because all members of an open cluster share the same age, combining the five per-star age estimates provides a stronger constraint than any one star alone. We model the stars with a common age, while each keeps its own mass, metallicity and equatorial rotation velocity. The cluster-age posterior is then
\begin{equation}
    p(\tau \mid \mathrm{all\ stars}) \;\propto\; \frac{\prod_{i=1}^{N} p_i(\tau)}{\pi(\tau)^{N-1}} ,
    \label{eq:age_product}
\end{equation}
where $N=5$ and $p_i(\tau)$ is the age posterior of star $i$ after integrating over its mass, metallicity and rotation velocity. Multiplying the five posteriors includes the age prior $\pi(\tau)$ five times, so we divide by $\pi(\tau)^{N-1}$ to count it only once. Since the prior, $\pi(\tau)$, is uniform over $[50,200]$\,Myr, this correction affects only the normalisation. To combine the posteriors, we first smoothed each star's age samples with a Gaussian kernel density estimate over this range. We then multiplied the five densities and normalised the result. Changing the kernel width by a factor of 2.5 moved the median age by less than $0.1$\,Myr and the interval limits by less than $0.2$\,Myr. This combination assumes independent errors between stars. This is an approximation because the stars share an emulator and occupy a similar region of parameter space, where their emulator residuals may be correlated. Figure~\ref{fig:cluster_age} shows this conditional cluster posterior alongside the individual ages.

The individual ages in Table~\ref{tab:pleiades_posteriors} are mutually consistent. Multiplying their age posteriors gives a cluster age of $113.3\pm4.5$\,Myr (statistical only; $2\sigma$ interval $104.1$--$122.1$\,Myr). An inverse-variance weighted mean of the five medians with symmetrised $1\sigma$ intervals gives $113.4\pm4.6$\,Myr (Fig.~\ref{fig:cluster_age_choices}), with consistency statistic $\chi^2/\nu=0.61/4=0.15$. Omitting each star in turn gives four-star product ages of $111.5$--$115.2$\,Myr, all within the five-star $1\sigma$ interval, so no single star drives the result.

Our age agrees with the lithium-depletion-boundary estimate of $112\pm5$\,Myr from \citet{dahm_2015}, who note that including magnetic activity in the boundary brown dwarfs would imply $\approx100$\,Myr. Relocating the boundary with \textit{Gaia} DR2 parallaxes, \citet{galindo-guil_lithium_2022} found $127.4^{+6.3}_{-10}$\,Myr using BT-Settl models; their six model sets span 117.6--155.5\,Myr for the same boundary. Our estimate is consistent with the lower end of that determination and lies at the young end of the 110--160\,Myr rotating-isochrone range \citep{gossage_pleiades_2018}, whose upper end is driven by the more extended turn-off in the near-infrared colour--magnitude diagram.

To assess sensitivity to the external metallicity constraint, we repeated all five fits with a uniform prior over $Z \in [0.001, 0.026]$. The emulator-error correlations were recomputed under this prior, retaining the adopted global diagonal error scales and all other fit settings. Combining the resulting marginal age posteriors with metallicity independent for each star gives $108.5^{+5.5}_{-5.7}$\,Myr (Fig.~\ref{fig:cluster_age_choices}). Relative to the adopted analysis, the median is $4.9$\,Myr younger and the 68 per cent interval is 24 per cent wider. The age and metallicity constraints therefore persist without the Soderblom prior within this setup (Table~\ref{tab:metallicity_prior_sensitivity}).

Three of the six stars analysed by \citet{fox_machado_seismological_2006} are in our sample: V624\,Tau, V647\,Tau and V1187\,Tau (HD\,23194). Their identifications were an output of the frequency matching rather than an input, and for four of their six stars more than one hundred identification combinations survived their selection criterion. Mode identification is the most plausible source of the difference between their age estimate and ours.

\subsection{Sensitivity to input physics}
\label{sec:input_physics_sensitivity}

The input physics adopted in the underlying \mesa grid \citep{gautam_modelling_2025} fixes the convective-core overshooting, mixing-length parameter and initial helium abundance. To quantify the sensitivity of the inferred ages to these choices, we generated synthetic stars from companion \storm grids in which one input parameter was varied at a time. We then fitted these stars using the emulator trained on the unperturbed grid. Each perturbed target was paired with an unperturbed control at the same mass, metallicity, initial rotation rate and age, using the same selected modes and observational-noise draw; the measured shift is the perturbed-minus-control difference in posterior-median age. Pairing removes emulator error common to the two fits. Each perturbation contains 100 pairs, giving 600 fits drawn from a subset of the Pleiades age domain. Appendix~\ref{app:physics} gives the detailed configuration and selection.

\newpage
\enlargethispage{4pt}
\begin{placedfigure}
    \centering
    \includegraphics[width=\columnwidth]{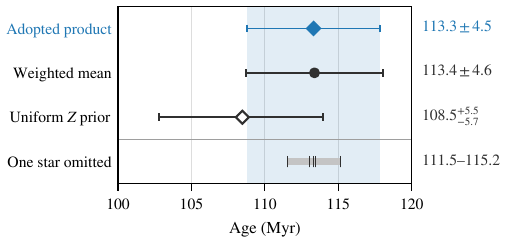}
    \caption{Sensitivity of the Pleiades cluster age to the combination method, the metallicity prior and the choice of stars. Diamonds mark medians of the product of the age posteriors of the five Pleiads (Eq.~\eqref{eq:age_product}), filled blue for the adopted fits and open for the fits repeated for a uniform prior over $Z \in [0.001, 0.026]$ with the emulator-error correlations recomputed under that prior. The circle marks the inverse-variance weighted mean of the five posterior medians with symmetrised $1\sigma$  intervals. Bars above the horizontal grey line show $1\sigma$  statistical intervals, and the shaded band marks the interval of the adopted product. Below the line, ticks mark the medians of the five four-star products, each omitting a different star. The grey bar spans their range; it is not an uncertainty interval. Omitting V624\,Tau gives the lowest median and omitting HD\,23325 the highest.}
    \label{fig:cluster_age_choices}
\end{placedfigure}

Changing $f_{\rm ov}$ by $\pm0.005$, $\alpha_{\rm MLT}$ by $\pm0.1$ and the initial helium mass fraction by $|\Delta Y|=0.02$ produces median absolute age shifts of $2.1$, $2.3$ and $5.2$\,Myr, respectively (Fig.~\ref{fig:physics_distributions}). The 16th--84th percentile ranges of the shifts are $0.6$--$5.2$, $0.2$--$6.6$ and $1.4$--$13.0$\,Myr, and the 90th percentiles are $6.5$, $7.6$ and $16.2$\,Myr. The median shifts for overshooting and mixing length are smaller than the $8$--$18$\,Myr statistical uncertainties on the individual stellar ages, although their upper tails can be comparable to these uncertainties. The median helium shift is comparable to the statistical uncertainty on the combined cluster age. For the perturbations tested here, these shifts indicate the scale of age errors that could arise if the model assumptions do not adequately describe the stars. Appendix~\ref{app:physics} gives the population coverage, mass dependence, boundary checks and signed parameter responses; Paper~III presents a more extensive treatment.

\newpage
\begin{placedfigure}
    \centering
    \includegraphics[width=\columnwidth]{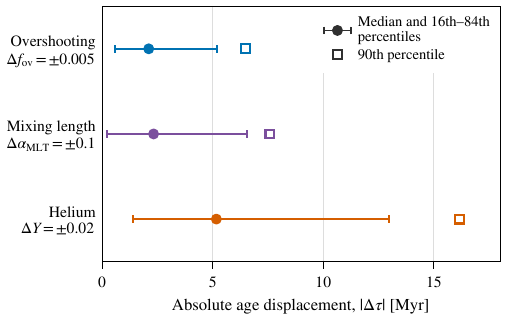}
    \caption{Input-physics age displacements for the interior-truth population. Filled circles show the median absolute difference between paired posterior-median ages, and bars span the 16th to 84th percentiles of these differences. Open squares show the 90th percentile. Each statistic summarises 100 perturbed/control pairs with shared true mass, metallicity, initial rotation rate and age, and shared selected modes and observational noise. The bars show how the difference varies from pair to pair. Table~\ref{tab:physics_population} gives 95 per cent bootstrap intervals for the 90th percentiles.}
    \label{fig:physics_distributions}
\end{placedfigure}
\setlength{\dbltextfloatsep}{15pt plus 4.5pt minus 3pt}

\section{Conclusions}
\label{sec:conclusions}

We have developed a framework for estimating the ages of young \dsct stars from individual pulsation frequencies, using a neural-network emulator to make Bayesian inference with rotating models practical. Trained on the \storm realisation of the Paper~I grid, the emulator predicts p\:mode, f\:mode and low-order g\:mode frequencies, including rotational splitting. Coupling these predictions to \ultranest allows mass, metallicity, age and equatorial rotation velocity to be inferred jointly, while propagating correlated emulator residuals through the likelihood.

We tested the framework using synthetic observations derived from two sources: emulator predictions and held-out grid models. Observations generated by the emulator are recovered with near-nominal coverage under matched simulation and inference assumptions, which supports the adopted likelihood, covariance construction and sampler. Tests using held-out grid models reveal a phase-dependent limitation: age credible intervals can become too narrow, with the largest recovery failures concentrated at $X_{\rm c}/X_0 \approx 0.77$--$0.91$. The five Pleiades stars lie outside this interval, at $X_{\rm c}/X_0 = 0.963$--$0.974$. 

Applied to five \dsct stars in the Pleiades observed by \textit{TESS}, the framework gives mutually consistent individual ages. Combining their marginal age posteriors gives $113.3\pm4.5$\,Myr (statistical), assuming a common age and independent errors between stars, with metallicity fitted separately for each star. This agrees with the $112\pm5$\,Myr lithium-depletion-boundary estimate, is consistent with the lower end of its \textit{Gaia}-relocated determination, and lies at the young end of the 110--160\,Myr rotating-isochrone range discussed in Section~\ref{sec:cluster_age}. 

The quoted statistical interval on the combined age does not include uncertainty in the adopted input physics or mode identifications. Separate tests using 100 perturbed/control pairs per input-physics variation, drawn from an interior subset of the $50$--$200$\,Myr age window, give median absolute age shifts of $2.1$, $2.3$ and $5.2$\,Myr for $|\Delta f_{\rm ov}|=0.005$, $|\Delta\alpha_{\rm MLT}|=0.1$ and $|\Delta Y|=0.02$, respectively. The corresponding 90th percentiles are $6.5$, $7.6$ and $16.2$\,Myr. For the perturbations and synthetic population tested here, these shifts indicate the sensitivity of the inferred ages to changes in the adopted input physics. They do not quantify a common bias for the five observed stars or define a probabilistic systematic uncertainty. Paper~III in this series extends the input-physics investigation.

Applying the framework to additional Pleiades members, and to \dsct pulsators in other young clusters and associations such as the Hyades, the Pisces--Eridanus stream and the Cepheus--Hercules complex, could tighten statistical constraints on cluster ages and allow the framework to be tested across stellar populations of different ages.

The same machinery could be applied to field \dsct stars, which have no cluster age to inherit and whose ages otherwise rest almost entirely on isochrone placement (Section~\ref{sec:intro}). Nothing in the likelihood requires cluster membership. An informative metallicity prior is not essential for obtaining constrained posteriors for the five targets tested here (Appendix~\ref{app:metallicity_prior}), but its importance for a field star would need to be assessed for that target. Field-star applications would also require reliable mode identifications, a broader age prior and separate calibration across the relevant evolutionary phases.

The Pleiades application demonstrates that neural-network emulation makes inference from individual pulsation frequencies practical for young \dsct stars. For the five stars studied here, the statistical age uncertainties are approximately 7--16 per cent individually and 4 per cent for the combined cluster age. These results support the development of asteroseismology as a complementary age estimation method for young intermediate-mass stars. The recovery tests and input-physics experiments identify limitations that need to be addressed when applying the method more widely.

\section*{Acknowledgements}
\phantomsection\label{sec:acknowledgements}

AG was supported by an Australian Government Research Training Program (RTP) Scholarship and a University of Southern Queensland (UniSQ) International Fees Research Scholarship. SJM was supported by the Australian Research Council (ARC) through Future Fellowship FT210100485. We thank Owen Scutt and Guy Davies (University of Birmingham) for many helpful discussions about neural-network emulators and Bayesian inference in asteroseismology. We also thank members of the USyd--UniSQ \dsct Research Group for ongoing feedback. This paper includes data collected by the \textit{TESS} mission. Funding for the \textit{TESS} mission is provided by the NASA Explorer Program.

\section*{Data availability}
\phantomsection\label{sec:data_availability}

The trained emulator and the inference code are available at the \href{https://github.com/gautam-404/velociscuti-inference}{velociscuti-inference} repository. The repository holds the network weights and the fitted input--output transforms, the per-output emulator uncertainties, the machine-readable mode identification tables adopted for the five Pleiades stars with their frequencies, amplitudes and identifications, the covariance matrices entering the likelihood, the posterior summaries reported here, and notebooks that carry out the five star fits and the cluster-age combination with the \ultranest settings of Section~\ref{sec:inference_sampler}. It also holds the held-out grid residuals from which the emulator covariance is estimated: the subset inside the priors adopted here, and the complete set of $2.26\times10^6$ models. An \href{https://gautam-404.github.io/velociscuti/}{interactive web demonstration} allows exploration of how predicted pulsation frequencies vary with stellar parameters. The underlying stellar pulsation grid is available at \href{https://zenodo.org/records/15839323}{Zenodo} and the corresponding \mesa and \gyre input files at the \href{https://github.com/gautam-404/dSct-model-grid}{dSct-model-grid} repository, as described in Paper~I \citep{gautam_modelling_2025}.

\subsection*{Software}

The stellar models and pulsation frequencies used in this work were produced with \mesa \citep{paxton_modules_2011, paxton_modules_2013, paxton_modules_2015, paxton_modules_2018, paxton_modules_2019, jermyn_modules_2023}, \gyre \citep{townsend_gyre_2013}, and \storm \citep{vanlaer_storm}. The analysis used Python and its libraries, such as NumPy, SciPy, pandas, PyArrow, scikit-learn, Astropy, Lightkurve, h5py, Matplotlib, seaborn, and corner. Neural-network training and inference used PyTorch, Lightning, and \ultranest \citep{buchner_ultranest_2021}. The iterative frequency extraction code is available at the \href{https://github.com/gautam-404/prewhitener}{prewhitener} repository.

\bibliographystyle{mnras}
\bibliography{bibliography, zotero_export}
\suppressfloats[t]

\appendix

\section{Mode identifications}
\label{app:modeids}

Tables~\ref{tab:modeids_v624}, \ref{tab:modeids_v647}, \ref{tab:modeids_v1187}, \ref{tab:modeids_hd23157} and \ref{tab:modeids_hd23325} list the pulsation modes of V624\,Tau, V647\,Tau, V1187\,Tau, HD\,23157 and HD\,23325 that enter the likelihood, with their measured frequencies, $1\sigma$  uncertainties, and adopted $(n, \ell, m)$ identifications. These are the revised identifications used for the posteriors reported in Section~\ref{sec:posteriors}. 

\begin{figure*}
    \centering
    \includegraphics[width=\textwidth]{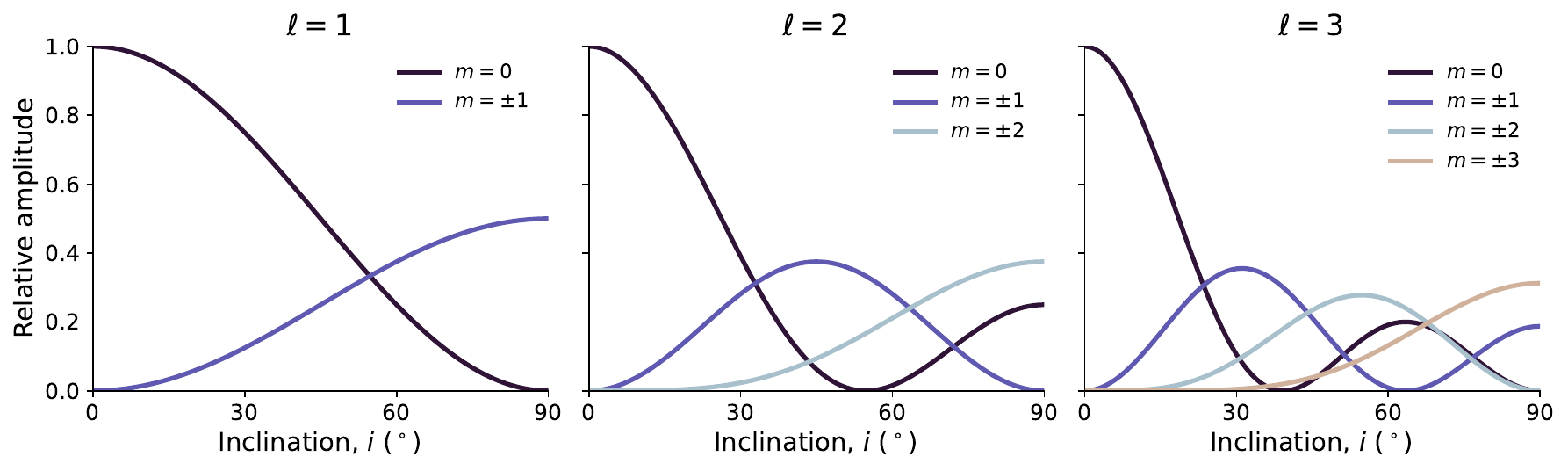}
    \caption{Geometric (disc-integrated) visibility of the rotationally split $m$-components of a non-radial mode as a function of stellar inclination $i$, for degrees $\ell = 1$, 2 and 3. Each curve is the relative visibility $V_{\ell m}(i) = \frac{(\ell-|m|)!}{(\ell+|m|)!}\,\bigl[P_\ell^{|m|}(\cos i)\bigr]^2$ (the power with which an $(\ell, m)$ component appears in disc-integrated light).}
    \label{fig:mode_visibility}
\end{figure*}

\begin{table}
    \centering
    \footnotesize
    \caption{Pulsation mode identifications for V624\,Tau. The mode type is p ($n>0$), f ($n=0$), or g ($n<0$, with $|n|$ the g\:mode radial order). Frequencies are from iterative prewhitening of the \textit{TESS} light curve. Peaks without a secure identification, and tentative identifications excluded from the fit, are not listed.}
    \label{tab:modeids_v624}
    \begin{tabular}{lr r c ccc}
        \hline
        ID & $\nu$\,(d$^{-1}$) & $\sigma_\nu$\,(d$^{-1}$) & Type & $n$ & $\ell$ & $m$ \\
        \hline
        F1 & 18.8138 & $4.0\times10^{-6}$ & g & -2 & 3 & 3 \\
        F2 & 20.9829 & $2.8\times10^{-6}$ & f & 0 & 2 & -2 \\
        F3 & 21.7921 & $3.4\times10^{-5}$ & p & 1 & 1 & 0 \\
        F4 & 22.4121 & $5.4\times10^{-5}$ & p & 1 & 1 & 1 \\
        F5 & 23.1360 & $1.3\times10^{-5}$ & f & 0 & 3 & -2 \\
        F6 & 23.8392 & $7.7\times10^{-6}$ & f & 0 & 3 & -1 \\
        F7 & 24.4054 & $8.1\times10^{-5}$ & f & 0 & 3 & 0 \\
        F8 & 25.0142 & $3.4\times10^{-5}$ & f & 0 & 3 & 1 \\
        F9 & 26.2144 & $7.8\times10^{-6}$ & f & 0 & 3 & 3 \\
        F10 & 27.4645 & $6.9\times10^{-6}$ & p & 2 & 0 & 0 \\
        F11 & 32.3818 & $1.0\times10^{-5}$ & p & 2 & 2 & 1 \\
        F12 & 32.9347 & $6.7\times10^{-6}$ & p & 2 & 2 & 2 \\
        F13 & 33.2308 & $4.0\times10^{-5}$ & p & 3 & 0 & 0 \\
        F14 & 33.9521 & $1.6\times10^{-5}$ & p & 2 & 3 & 0 \\
        F15 & 35.3399 & $4.0\times10^{-6}$ & p & 3 & 1 & 0 \\
        F16 & 35.7145 & $2.3\times10^{-6}$ & p & 2 & 3 & 3 \\
        F17 & 35.9687 & $6.5\times10^{-6}$ & p & 3 & 1 & 1 \\
        F18 & 41.2826 & $3.8\times10^{-5}$ & p & 3 & 3 & 1 \\
        F19 & 42.2785 & $4.9\times10^{-6}$ & p & 4 & 1 & 0 \\
        F20 & 42.8191 & $5.6\times10^{-5}$ & p & 4 & 1 & 1 \\
        F21 & 45.7048 & $3.5\times10^{-6}$ & p & 5 & 0 & 0 \\
        F22 & 49.2945 & $2.6\times10^{-5}$ & p & 5 & 1 & 0 \\
        F23 & 52.6961 & $2.7\times10^{-5}$ & p & 6 & 0 & 0 \\
        F24 & 56.3114 & $1.4\times10^{-5}$ & p & 6 & 1 & 0 \\
        F25 & 59.6847 & $2.0\times10^{-5}$ & p & 7 & 0 & 0 \\
        F26 & 63.2327 & $5.3\times10^{-5}$ & p & 7 & 1 & 0 \\
        \hline
    \end{tabular}
\end{table}

\begin{table}
    \centering
    \footnotesize
    \caption{Pulsation mode identifications for V647\,Tau. Columns and conventions as in Table~\ref{tab:modeids_v624}.}
    \label{tab:modeids_v647}
    \begin{tabular}{lr r c ccc}
        \hline
        ID & $\nu$\,(d$^{-1}$) & $\sigma_\nu$\,(d$^{-1}$) & Type & $n$ & $\ell$ & $m$ \\
        \hline
        F1 & 17.8898 & $8.2\times10^{-4}$ & g & -1 & 2 & 0 \\
        F2 & 18.6883 & $7.6\times10^{-5}$ & g & -1 & 2 & 2 \\
        F3 & 20.8493 & $2.8\times10^{-4}$ & p & 1 & 0 & 0 \\
        F4 & 21.3879 & $7.7\times10^{-4}$ & p & 1 & 1 & 0 \\
        F5 & 21.9161 & $2.2\times10^{-4}$ & p & 1 & 1 & 1 \\
        F6 & 23.0985 & $9.5\times10^{-4}$ & f & 0 & 3 & -2 \\
        F7 & 23.6541 & $2.8\times10^{-4}$ & f & 0 & 3 & -1 \\
        F8 & 25.1444 & $9.8\times10^{-4}$ & f & 0 & 3 & 2 \\
        F9 & 26.3238 & $5.0\times10^{-5}$ & p & 1 & 2 & 0 \\
        F10 & 26.7416 & $2.6\times10^{-4}$ & p & 2 & 0 & 0 \\
        F11 & 27.2628 & $6.3\times10^{-5}$ & p & 1 & 2 & 2 \\
        F12 & 28.4760 & $2.4\times10^{-4}$ & p & 2 & 1 & 1 \\
        F13 & 35.0458 & $9.7\times10^{-5}$ & p & 3 & 1 & 1 \\
        F14 & 41.4605 & $1.8\times10^{-3}$ & p & 4 & 1 & 0 \\
        F15 & 41.9030 & $1.3\times10^{-3}$ & p & 4 & 1 & 1 \\
        F16 & 44.9564 & $6.5\times10^{-4}$ & p & 5 & 0 & 0 \\
        F17 & 48.8321 & $8.3\times10^{-4}$ & p & 5 & 1 & 1 \\
        F18 & 51.8281 & $1.6\times10^{-3}$ & p & 6 & 0 & 0 \\
        F19 & 55.2353 & $1.5\times10^{-3}$ & p & 6 & 1 & 0 \\
        F20 & 58.7158 & $5.0\times10^{-4}$ & p & 7 & 0 & 0 \\
        \hline
    \end{tabular}
\end{table}

\begin{table}
    \centering
    \footnotesize
    \caption{Pulsation mode identifications for V1187\,Tau. Columns and conventions as in Table~\ref{tab:modeids_v624}.}
    \label{tab:modeids_v1187}
    \begin{tabular}{lr r c ccc}
        \hline
        ID & $\nu$\,(d$^{-1}$) & $\sigma_\nu$\,(d$^{-1}$) & Type & $n$ & $\ell$ & $m$ \\
        \hline
        F1 & 19.8146 & $5.4\times10^{-5}$ & g & -1 & 3 & -2 \\
        F2 & 20.2863 & $9.8\times10^{-5}$ & g & -1 & 3 & -1 \\
        F3 & 20.8101 & $5.6\times10^{-5}$ & g & -1 & 3 & 0 \\
        F4 & 22.3774 & $1.3\times10^{-4}$ & f & 0 & 2 & 0 \\
        F5 & 22.8867 & $5.8\times10^{-5}$ & f & 0 & 2 & 1 \\
        F6 & 23.4807 & $2.2\times10^{-5}$ & f & 0 & 2 & 2 \\
        F7 & 28.3198 & $3.2\times10^{-5}$ & p & 1 & 3 & 0 \\
        F8 & 28.7418 & $1.3\times10^{-4}$ & p & 2 & 1 & 0 \\
        F9 & 29.8289 & $1.1\times10^{-4}$ & p & 1 & 3 & 3 \\
        F10 & 35.5567 & $5.5\times10^{-5}$ & p & 3 & 1 & 0 \\
        F11 & 39.5199 & $3.9\times10^{-5}$ & p & 4 & 0 & 0 \\
        F12 & 42.6028 & $2.3\times10^{-5}$ & p & 4 & 1 & 0 \\
        F13 & 42.9544 & $1.4\times10^{-5}$ & p & 4 & 1 & 1 \\
        F14 & 46.1052 & $1.1\times10^{-6}$ & p & 5 & 0 & 0 \\
        F15 & 49.6752 & $1.5\times10^{-6}$ & p & 5 & 1 & 0 \\
        F16 & 50.0748 & $8.8\times10^{-5}$ & p & 5 & 1 & 1 \\
        F17 & 53.1756 & $1.2\times10^{-6}$ & p & 6 & 0 & 0 \\
        F18 & 56.7051 & $9.7\times10^{-5}$ & p & 6 & 1 & 0 \\
        F19 & 60.1274 & $1.4\times10^{-4}$ & p & 7 & 0 & 0 \\
        F20 & 63.6272 & $4.5\times10^{-5}$ & p & 7 & 1 & 0 \\
        F21 & 64.1060 & $6.0\times10^{-5}$ & p & 7 & 1 & 1 \\
        F22 & 71.0564 & $7.6\times10^{-5}$ & p & 8 & 1 & 1 \\
        \hline
    \end{tabular}
\end{table}

\begin{table}
    \centering
    \footnotesize
    \caption{Pulsation mode identifications for HD\,23157. Columns and conventions as in Table~\ref{tab:modeids_v624}. Rows F1--F19 enter the likelihood. Rows X1--X3 carry tentative identifications but are excluded from the fit on the criterion of Section~\ref{sec:modeid}.}
    \label{tab:modeids_hd23157}
    \begin{tabular}{lr r c ccc}
        \hline
        ID & $\nu$\,(d$^{-1}$) & $\sigma_\nu$\,(d$^{-1}$) & Type & $n$ & $\ell$ & $m$ \\
        \hline
        F1 & 17.0396 & $4.6\times10^{-6}$ & g & -1 & 2 & -2 \\
        F2 & 18.0193 & $1.5\times10^{-5}$ & g & -1 & 3 & -3 \\
        F3 & 19.8487 & $3.7\times10^{-5}$ & g & -1 & 3 & -1 \\
        F4 & 20.2604 & $2.0\times10^{-5}$ & g & -1 & 2 & 2 \\
        F5 & 20.9993 & $7.9\times10^{-5}$ & p & 1 & 0 & 0 \\
        F6 & 21.7700 & $6.8\times10^{-6}$ & g & -1 & 3 & 1 \\
        F7 & 23.7355 & $6.8\times10^{-6}$ & g & -1 & 3 & 3 \\
        F8 & 30.7211 & $5.7\times10^{-5}$ & p & 2 & 2 & -1 \\
        F9 & 31.1198 & $9.4\times10^{-5}$ & p & 1 & 3 & 3 \\
        F10 & 31.6371 & $8.2\times10^{-6}$ & p & 2 & 2 & 0 \\
        F11 & 32.4577 & $2.3\times10^{-6}$ & p & 2 & 2 & 1 \\
        F12 & 32.6255 & $5.7\times10^{-5}$ & p & 2 & 3 & -1 \\
        F13 & 34.5188 & $8.3\times10^{-5}$ & p & 2 & 3 & 1 \\
        F14 & 35.4320 & $6.6\times10^{-6}$ & p & 2 & 3 & 2 \\
        F15 & 38.1879 & $2.1\times10^{-5}$ & p & 3 & 3 & -2 \\
        F16 & 38.5349 & $4.7\times10^{-6}$ & p & 3 & 2 & 1 \\
        F17 & 41.5404 & $6.1\times10^{-5}$ & p & 4 & 1 & 0 \\
        F18 & 45.2712 & $9.3\times10^{-5}$ & p & 5 & 0 & 0 \\
        F19 & 55.2495 & $8.1\times10^{-5}$ & p & 6 & 1 & 0 \\
        \hline
        X1 & 18.6872 & $2.9\times10^{-5}$ & g & -1 & 2 & 0 \\
        X2 & 27.0943 & $4.1\times10^{-5}$ & p & 2 & 0 & 0 \\
        X3 & 52.0799 & $2.0\times10^{-5}$ & p & 6 & 0 & 0 \\
        \hline
    \end{tabular}
\end{table}

\begin{table}
    \centering
    \footnotesize
    \caption{Pulsation mode identifications for HD\,23325. Columns and conventions as in Table~\ref{tab:modeids_v624}.}
    \label{tab:modeids_hd23325}
    \begin{tabular}{lr r c ccc}
        \hline
        ID & $\nu$\,(d$^{-1}$) & $\sigma_\nu$\,(d$^{-1}$) & Type & $n$ & $\ell$ & $m$ \\
        \hline
        F1 & 15.5820 & $8.2\times10^{-5}$ & g & -2 & 2 & 2 \\
        F2 & 16.4889 & $4.5\times10^{-5}$ & g & -1 & 2 & -2 \\
        F3 & 18.2959 & $1.7\times10^{-5}$ & g & -1 & 2 & 0 \\
        F4 & 21.2336 & $5.3\times10^{-5}$ & p & 1 & 0 & 0 \\
        F5 & 21.7128 & $8.4\times10^{-6}$ & p & 1 & 1 & 0 \\
        F6 & 28.2180 & $2.3\times10^{-5}$ & p & 2 & 1 & 0 \\
        F7 & 29.5025 & $6.2\times10^{-6}$ & p & 2 & 1 & 1 \\
        F8 & 36.1510 & $4.7\times10^{-6}$ & p & 3 & 1 & 1 \\
        F9 & 38.9946 & $2.0\times10^{-5}$ & p & 4 & 0 & 0 \\
        F10 & 41.8485 & $1.9\times10^{-5}$ & p & 4 & 1 & 0 \\
        F11 & 43.0157 & $8.0\times10^{-5}$ & p & 4 & 1 & 1 \\
        F12 & 45.2805 & $6.1\times10^{-5}$ & p & 5 & 0 & 0 \\
        F13 & 49.8427 & $4.9\times10^{-5}$ & p & 5 & 1 & 1 \\
        F14 & 52.2196 & $2.1\times10^{-5}$ & p & 6 & 0 & 0 \\
        F15 & 59.0195 & $1.0\times10^{-4}$ & p & 7 & 0 & 0 \\
        F16 & 65.8040 & $6.8\times10^{-5}$ & p & 8 & 0 & 0 \\
        \hline
    \end{tabular}
\end{table}

\section{Inference convergence diagnostics}
\label{app:convergence}

Table~\ref{tab:convergence} lists the sampler diagnostics for the five adopted
observed-star runs. Each used at most $1.1\times10^{5}$ of the $10^{7}$ permitted
likelihood evaluations and returned an effective posterior sample size of
7\,739--11\,378 against the 1000 requested (Section~\ref{sec:inference_sampler}).
The runs stopped on logged evidence error-budget estimates of $0.06$--$0.07$,
within the requested \texttt{dlogz}$=0.1$, whereas the returned evidence
uncertainties are $0.10$--$0.15$.

The posterior mass at the mass and age prior boundaries is negligible for every
star, and no metallicity posterior approaches the grid boundary at $Z = 0.026$.
All five metallicity posteriors do, however, sit $1.3$--$2.1\sigma$ above the mean
of the truncated cluster prior, with V647\,Tau ($2.1\sigma$) and HD\,23157
($1.9\sigma$) the furthest into the upper tail.

\begin{table}
    \centering
    \caption{Convergence diagnostics of the five adopted runs: total likelihood evaluations $N_{\rm call}$, sampler iterations $N_{\rm iter}$, and the effective posterior sample size $N_{\rm eff}$.}
    \label{tab:convergence}
    \begin{tabular}{lrrr}
        \hline
        Star & $N_{\rm call}$ & $N_{\rm iter}$ & $N_{\rm eff}$\\
        \hline
        V624\,Tau  & 100\,229 & 43\,021 & 9\,918\\
        V647\,Tau  & 113\,974 & 36\,269 & 8\,247\\
        V1187\,Tau & 85\,433  & 29\,688 & 7\,739\\
        HD\,23157  & 113\,916 & 49\,009 & 11\,378\\
        HD\,23325  & 109\,110  & 42\,540 & 10\,435\\
        \hline
    \end{tabular}
\end{table}

\section{Evolutionary dependence of recovery failures}
\label{app:recovery_failures}

\begin{figure*}
    \centering
    \includegraphics[width=\textwidth]{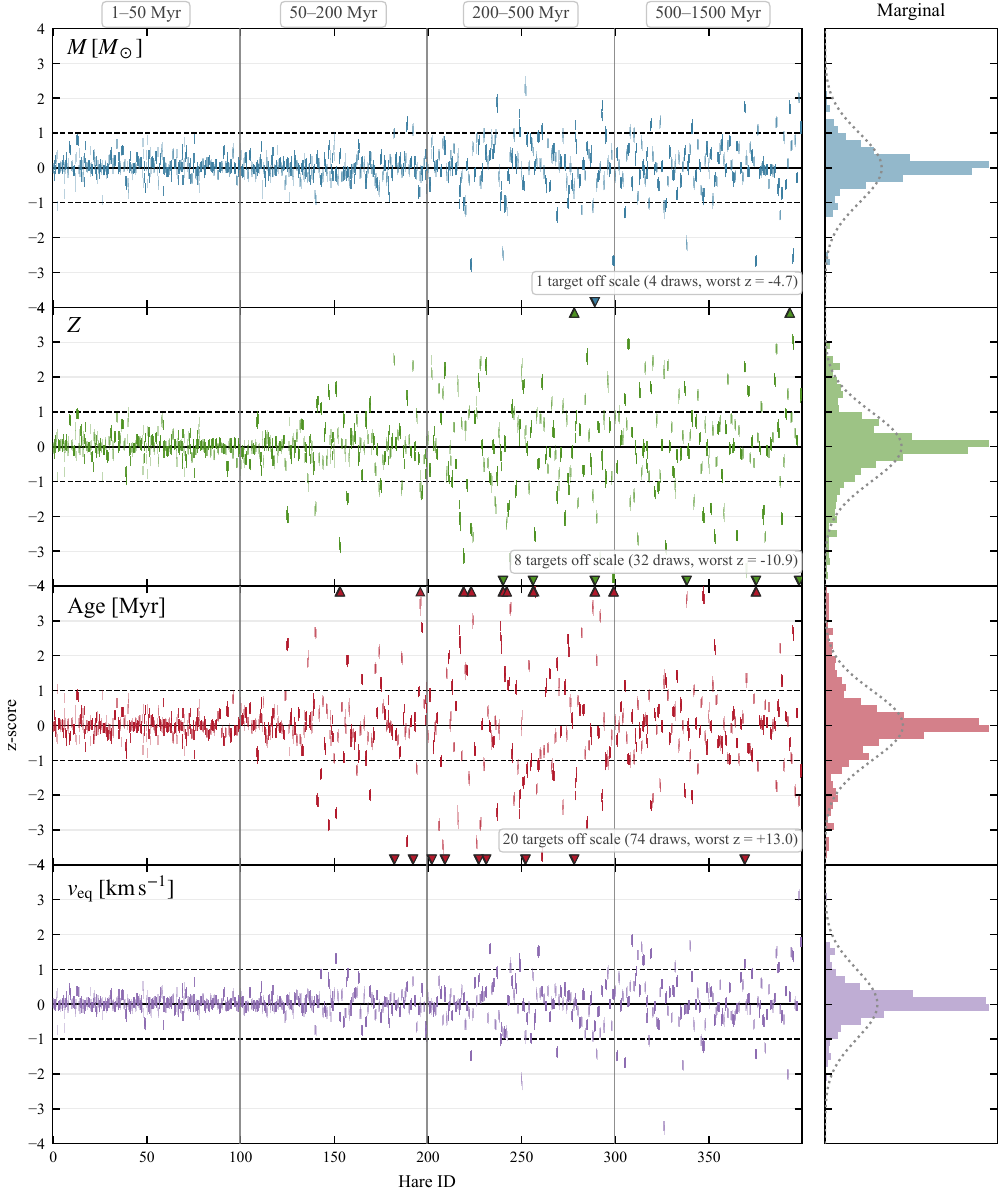}
    \caption{Standardised residuals for hares and noisy hares, with one row per parameter. Each tick shows one fit. Each target's four fits (one noiseless and three noisy) share the same horizontal position, with alternating shades distinguishing neighbouring targets. Targets are ordered by true age. Solid and dashed lines mark $z=0$ and $z=\pm1$. Residuals beyond $|z|=4$ are marked at the upper or lower edge and counted in each panel. Histograms on the right show the residual distributions, with dotted curves showing a standard normal distribution.}
    \label{fig:hounds_zscore_draws}
\end{figure*}

\begin{figure*}
    \centering
    \includegraphics[width=\textwidth]{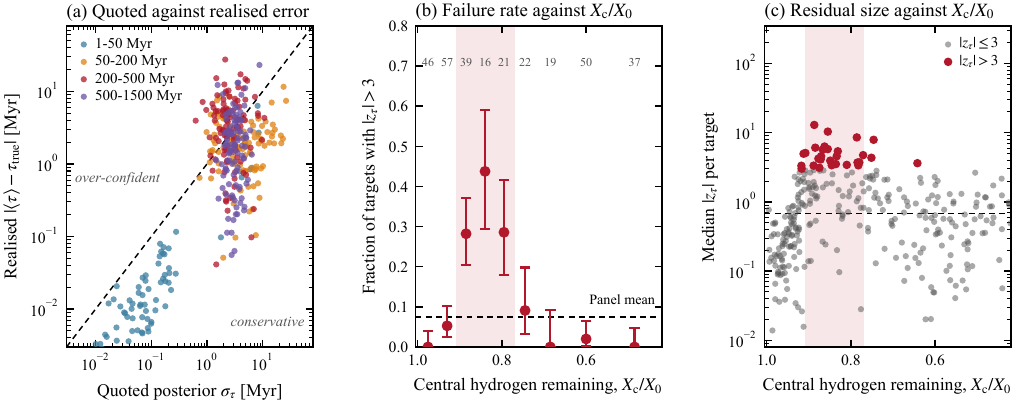}
    \caption{Evolutionary dependence of age recovery for hares and noisy hares. \emph{(a)} Median absolute age error against median posterior standard deviation, coloured by age band. \emph{(b)} Fraction of main-sequence targets with $|z_\tau|>3$ versus the fraction of initial central hydrogen remaining, $X_{\rm c}/X_0$. Error bars show $1\sigma$ intervals, numbers above the points give the targets per bin, and the dashed line marks the mean ($0.076$). \emph{(c)} Median $|z_\tau|$ versus $X_{\rm c}/X_0$. The dashed line at $0.674$ is the median $|z|$ for a standard normal distribution. Panels (a) and (c) show one point per target, with medians taken over its four fits. Shading marks the failure window, $0.77\leq X_{\rm c}/X_0\leq0.91$. Evolution proceeds from high to low $X_{\rm c}/X_0$ in (b) and (c).}
    \label{fig:hounds_failure_locus}
\end{figure*}

The target-level diagnostics in Figs.~\ref{fig:hounds_zscore_draws} and~\ref{fig:hounds_failure_locus} show that the age-recovery failures summarised in Section~\ref{sec:hh_global} concentrate at $0.77\leq X_{\rm c}/X_0\leq0.91$. This window contains 19 per cent of the targets and 80 per cent of the 30 age outliers. Of these outliers, 29 already exceed $|z_\tau|=3$ without observational noise and remain outliers in all three noisy fits. The hare and noisy-hare age-residual standard deviations agree to two decimal places.

Figure~\ref{fig:hounds_failure_locus} shows this dependence on evolutionary phase. Above $X_{\rm c}/X_0\approx0.95$, effectively all 139 targets recover the true age within $1\sigma$, with no age outliers. This coverage falls to $0.27$ in the failure window, where the largest standardised residuals occur near $X_{\rm c}/X_0\approx0.85$. Below $X_{\rm c}/X_0\approx0.65$, coverage is $0.78$, although some failures persist.

The emulator errors also vary with phase. The g and f\:mode errors peak near $X_{\rm c}/X_0=0.89$ and $0.82$, respectively, at roughly $\hat{\sigma}_{\rm model}$, while p-mode errors remain near $0.07\,\hat{\sigma}_{\rm model}$. The later p-mode peak, near $X_{\rm c}/X_0=0.5$, accounts for the residual failures at more evolved phases. The failure window coincides with avoided crossings between the low-order g and f\:modes and p\:modes. Along the grid tracks, we compare modes of the same spherical degree $\ell$. For both $\ell=2$ and $\ell=3$, the frequency gaps between the $n_{\rm pg}=-1$ g\:mode and the f\:mode, and between the f\:mode and the $n=1$ p\:mode, reach local minima of $0.38$--$0.45\,\Delta\nu$ at $X_{\rm c}/X_0=0.75$--$0.82$. These gaps widen again on either side of the minima. For $\ell=1$, the gap between the $n_{\rm pg}=-1$ g\:mode and the $n=1$ p\:mode reaches its minimum later in evolution, at $X_{\rm c}/X_0=0.68$--$0.70$. This is consistent with higher-degree g\:modes reaching the p\:mode frequency range earlier \citep{gautam_modelling_2025}. We therefore attribute the localised miscalibration to mixed-mode behaviour at these crossings. It is correctable in principle, but no correction is attempted here.

Age and metallicity errors are anti-correlated, consistent with the $\tau$--$Z$ degeneracy. For the 30 selected outliers, noiseless-hare posterior means give 17 older, more metal-poor solutions and 13 younger, more metal-rich solutions.

The 30 outliers use 28 different sets of fitted modes. Each of these mode sets is also used for two or three successfully recovered targets. Within the failure window, both outliers and recovered targets contain f and g modes, with no detectable differences in median f- and g-mode frequencies relative to $\Delta\nu$, frequency uncertainties or injected perturbations. 

Together with the noiseless failures, these results show that injected noise, any individual mode or any particular set of fitted modes is insufficient to explain the failures; evolutionary phase remains the dominant shared property. The physical age errors remain modest despite the large standardised residuals. Table~\ref{tab:hounds_age_errors} gives the age-band breakdown. After averaging each target's four posterior means, all 30 selected outliers lie within $6.7$ per cent of their true ages.

\begin{table*}
    \centering
    \caption{Absolute age errors after averaging each target's one noiseless-hare and three noisy-hare posterior means, $|\Delta\tau|=|\tau_{\rm inferred}-\tau_{\rm true}|$. Errors are in Myr except for the final column, which gives the relative age error ranges for the outliers.}
    \label{tab:hounds_age_errors}
    \begin{tabular}{lcccc}
        \hline
        \shortstack{Age band\\(Myr)} &
        \shortstack{All targets\\median (90th percentile)} &
        \shortstack{Number of\\age outliers} &
        \shortstack{Age outliers\\median (full range)} &
        \shortstack{Relative age error\\range (per cent)} \\
        \hline
        $1$--$50$     & $0.005$ ($0.061$) & 0  & -- & -- \\
        $50$--$200$   & $1.9$ ($4.5$)     & 5  & $7.54$ ($4.80$--$8.52$)   & $2.72$--$6.69$ \\
        $200$--$500$  & $3.5$ ($9.6$)     & 21 & $8.52$ ($3.95$--$27.50$)  & $1.44$--$6.38$ \\
        $500$--$1500$ & $1.7$ ($6.0$)     & 4  & $12.71$ ($9.15$--$13.19$) & $1.07$--$2.02$ \\
        \hline
    \end{tabular}
\end{table*}

\section{Input-physics displacement measurements}
\label{app:physics}

This appendix gives the configuration, target selection and calibration checks behind Section~\ref{sec:input_physics_sensitivity}. Within a pair the mass, metallicity, age, initial rotation rate, selected modes and observational-noise draw are identical, and only the evolved equatorial velocity can differ. Pairing cancels the error common to the two fits, while any differential emulator or fitting error remains.

Each parameter was tested with 100 pairs, each assigned a positive or negative perturbation and one shared noise realisation. Both members satisfy $1.5\leq M/{\rm M}_\odot\leq2.3$, $0.005\leq Z\leq0.022$ and $20\leq v_{\rm eq}\leq120$\,km\,s$^{-1}$. The sampled metallicities are $Z=0.005,0.010,0.015,0.020$, and the true ages span $65$--$185$\,Myr. These cuts keep the synthetic inputs inside the model domain; the fits retain the broader priors specified below.

The reference grid uses $f_{\rm ov}=0.017$, $\alpha_{\rm MLT}=1.9$ and $Y_{\rm ref}(Z)=0.28+1.4(Z-0.0142)$, rounded to four decimal places. Here, $0.28$ is the adopted initial-helium mass fraction at the adopted bulk-solar metallicity $Z_{\odot,{\rm bulk}}=0.0142$ \citep{asplund_chemical_2009}. The modified grids vary one parameter by the amounts in Table~\ref{tab:physics_population}.

The fits used the emulator of Section~\ref{sec:nn} with the same per-mode error scales $\hat{\sigma}_{\rm model}$ as the observed-star fits, and the inference framework of Section~\ref{sec:inference}. The frequency-error covariance was estimated from $10^6$ emulator samples using test and validation residuals within the prior limits, with shrinkage fixed at $0.1$ and no added frequency-error floor. Each fit used 1000 live points, a stopping tolerance $\Delta\ln Z=0.1$ and at most $10^7$ likelihood evaluations. All 600 fits satisfied the numerical convergence criteria without reaching this limit.

The priors are uniform over $50$--$200$\,Myr in age, $1.4$--$2.5$\,M$_\odot$ in mass and $0$--$200$\,km\,s$^{-1}$ in equatorial velocity. The metallicity prior is Gaussian, centred on the true $Z$, with $\sigma_Z=Z\ln(10)\sqrt{0.02^2+0.05^2}$ and truncated to $0.001\leq Z\leq0.026$. The covariance estimation and metallicity priors are therefore specified separately from those of the five observed-star fits.

\subsection{Parameter shifts and age coverage}

Table~\ref{tab:physics_population} reports absolute age displacements, $|\Delta\tau|$. Their 90th percentiles are $3.1$--$3.3$ times their medians (Fig.~\ref{fig:physics_distributions}). Age shifts can have either sign even for a fixed perturbation sign. These statistics depend on the selected stars and mode sets; they do not define a common age correction or a systematic uncertainty for the five Pleiades stars.

Mass and metallicity move predominantly in one direction once the sign of the helium offset is fixed, with the signed displacement taken between posterior means. At $\Delta Y=+0.02$, mass increases in 44 of 49 pairs and metallicity decreases in 44 of 49; at $\Delta Y=-0.02$, mass decreases in 44 of 51 and metallicity increases in 50 of 51. The positive-offset median displacements are $+0.0258$\,M$_\odot$ and $-2.99\times10^{-4}$ in $Z$. Dividing each pair's signed displacement by the perturbed fit's posterior standard deviation gives median ratios of $+5.1$ for mass and $-1.6$ for metallicity. These ratios measure displacement against the fitted width; they are not significances from a calibrated sampling distribution.

Across the three perturbations, the central 68 and 95 per cent age intervals of the control fits contain the injected age in 249 and 270 of 300 cases. Those cases sample 117 distinct evolutionary tracks, some appearing in more than one perturbation, and resampling whole tracks rather than individual fits gives 95 per cent intervals of $77.7$--$87.9$ and $85.9$--$93.7$ per cent for the two fractions. The 68 per cent intervals therefore contain the truth more often than they should, while the 95 per cent intervals still contain it too rarely.

Coverage also varies with mass. Both interval levels contain the truth in all 114 controls with $1.50\leq M/{\rm M}_\odot\leq1.75$, compared with 48 and 61 of 85 controls above $2.00$\,M$_\odot$. The median age posterior standard deviation falls from 9.16 to 1.96\,Myr between these subsets. These subsets were chosen after the fits and are descriptive; they do not provide a separate calibration of the five Pleiades stars. Table~\ref{tab:physics_population} gives coverage for each perturbation.

\begin{figure*}
    \centering
    \includegraphics[width=\textwidth]{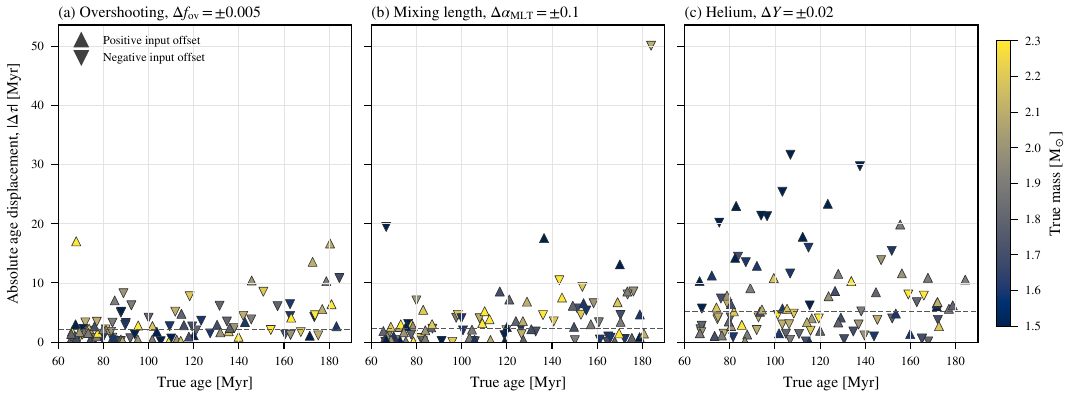}
    \caption{Absolute paired age displacement versus injected age for each input-physics perturbation. All 100 pairs per panel are shown. Colour denotes true mass; upward and downward triangles distinguish positive and negative input-physics offsets. Dashed horizontal lines mark the population median in each panel. Axes are shared across panels, and no outliers are removed.}
    \label{fig:physics_truth_dependence}
\end{figure*}

\begin{figure}
    \centering
    \includegraphics[width=0.98\columnwidth]{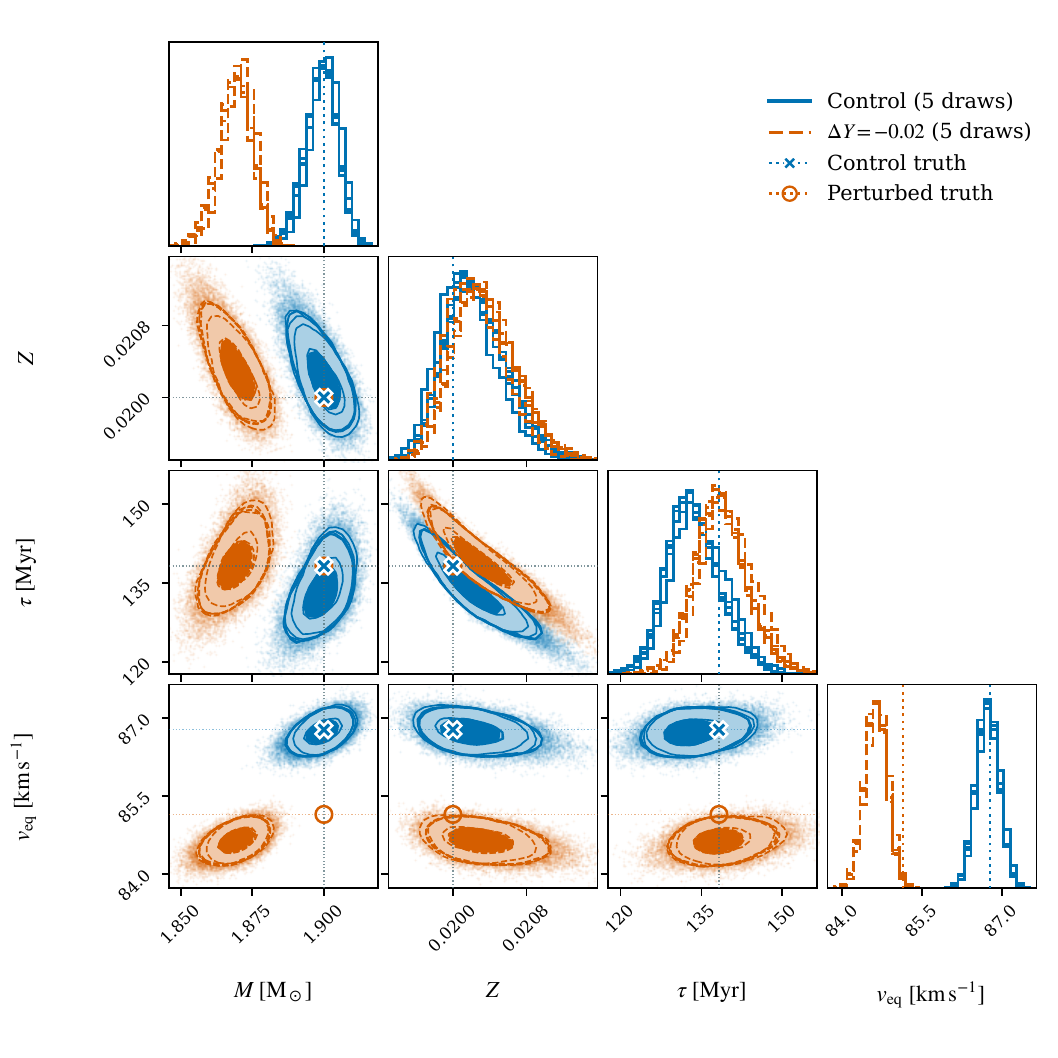}
    \caption{Helium-response illustration selected near the median of the original 100-pair $|\Delta Y|=0.02$ population. This pair has a displacement of $5.190$\,Myr, compared with the population median of $5.166$\,Myr. Its truth parameters are $M=1.90$\,M$_\odot$, $Z=0.020$ and $\tau=138.266$\,Myr; the evolved equatorial velocities are $86.776$\,km\,s$^{-1}$ for the control and $85.145$\,km\,s$^{-1}$ for $\Delta Y=-0.02$. Blue solid curves show five control fits and orange dashed curves their five paired perturbed fits. The paired posterior-mean age differences span $5.066$--$5.180$\,Myr. Crosses and open circles mark the control and perturbed truths, which coincide in mass, metallicity and age.}
    \label{fig:physics_helium_median}
\end{figure}

\begin{figure}
    \centering
    \includegraphics[width=0.98\columnwidth]{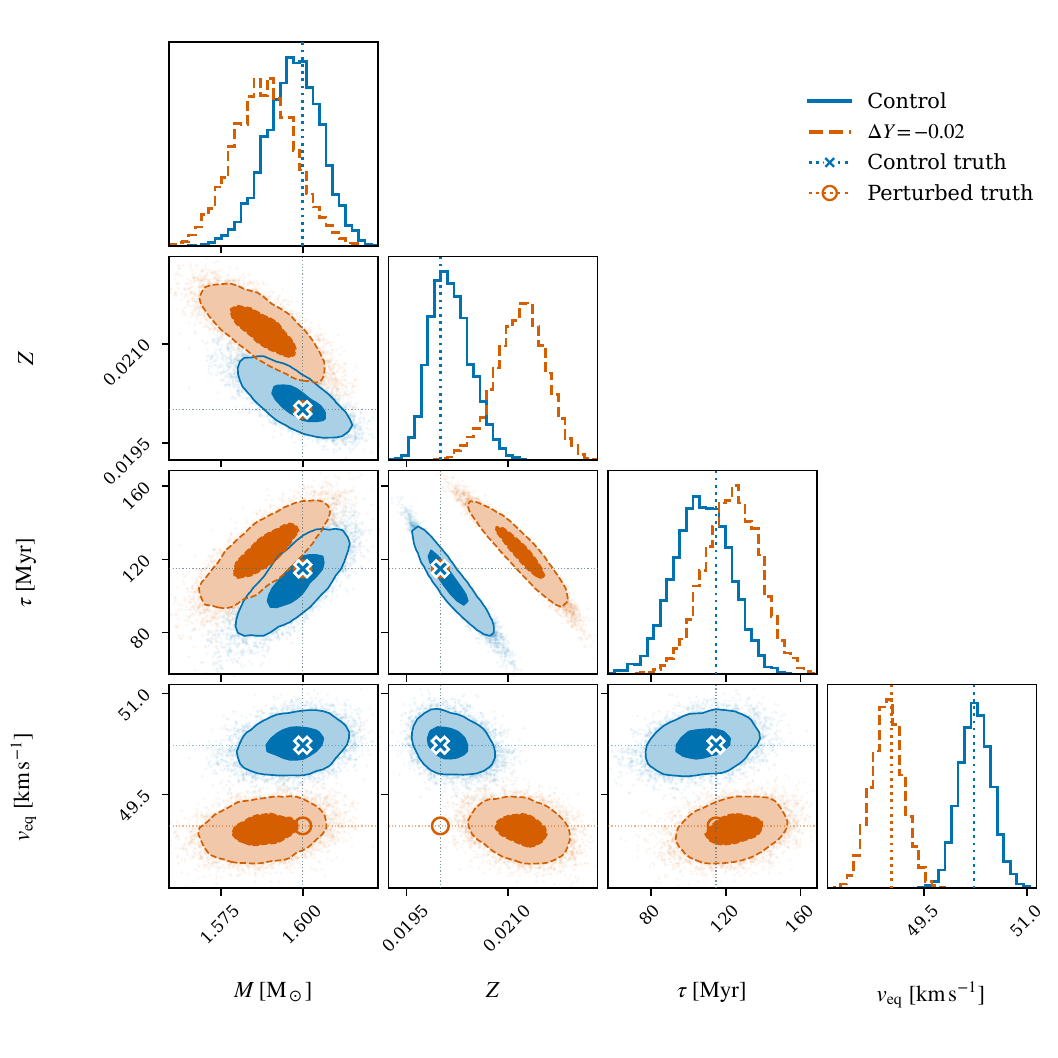}
    \caption{Upper-tail helium-response illustration for the eligible pair closest to the linearly interpolated 90th percentile of the original 100-pair $|\Delta Y|=0.02$ population. This pair's posterior median age displacement is $15.972$\,Myr, compared with the population 90th percentile of $16.156$\,Myr. The original control and perturbed fits have truths $M=1.60$\,M$_\odot$, $Z=0.020$ and $\tau=114.873$\,Myr, with evolved equatorial velocities of $50.232$ and $49.030$\,km\,s$^{-1}$, respectively. Blue solid and orange dashed curves show the control and $\Delta Y=-0.02$ fits. Crosses and open circles mark the control and perturbed truths, which coincide in mass, metallicity and age.}
    \label{fig:physics_helium_p90}
\end{figure}

Figure~\ref{fig:physics_truth_dependence} shows how the age displacements vary across the sample, and Figs.~\ref{fig:physics_helium_median} and \ref{fig:physics_helium_p90} show single $|\Delta Y|=0.02$ pairs selected near the population median and 90th percentile. The two pairs illustrate those responses; the quantiles themselves come from all 100 helium pairs.

\begin{table*}
    \centering
    \caption{Input-physics age displacements in the interior-truth population: 600 fits, with one observational-noise draw per perturbed/control pair. The median and 90th percentile summarise the absolute difference in posterior-median age over $N$ pairs. Values in parentheses are 95 per cent bootstrap intervals of the 90th percentile from 20\,000 resamplings of the $N$ pairs with replacement. Coverage entries give the number of central 68 and 95 per cent age intervals containing the injected truth, each out of 100.}
    \label{tab:physics_population}
    \begin{tabular}{llrrrcc}
        \hline
        Perturbation & Size & $N$ pairs & median $|\Delta \tau|$ [Myr] & $p_{90}$ [Myr] & Control 68/95 & Perturbed 68/95 \\
        \hline
        Convective-core overshooting & $\Delta f_{\rm ov}=\pm0.005$ & 100 & 2.1 & 6.5 (5.1--10.3) & 84/90 & 80/86 \\
        Mixing length & $\Delta\alpha_{\rm MLT}=\pm0.1$ & 100 & 2.3 & 7.6 (6.2--9.3) & 79/87 & 72/83 \\
        Initial helium & $\Delta Y=\pm0.02$ & 100 & 5.2 & 16.2 (11.7--21.4) & 86/93 & 56/79 \\
        \hline
    \end{tabular}
\end{table*}

\section{Sensitivity to the metallicity prior}
\label{app:metallicity_prior}

Table~\ref{tab:metallicity_prior_sensitivity} compares all four fitted parameters under the adopted Soderblom metallicity prior and a uniform prior over $Z \in [0.001, 0.026]$. Both cases retain the $[50, 200]$\,Myr age prior. For the uniform prior fits, we recomputed the emulator error correlations while retaining the global diagonal error scales and all other fit settings. The cluster row gives the product of the five marginal age posteriors evaluated as in Section~\ref{sec:cluster_age}, with metallicity independent for each star.

\begin{table*}
    \centering
    \caption{Sensitivity of the five-star inference to the metallicity prior. In the two fit groups, entries are posterior medians with asymmetric $1\sigma$ (16th--84th-percentile) uncertainties; the metallicity columns give $10^3 Z$. The last four columns give each flat-prior median minus the corresponding Soderblom-prior median, computed from unrounded values.}
    \label{tab:metallicity_prior_sensitivity}
    \small
    \setlength{\tabcolsep}{3.5pt}
    \renewcommand{\arraystretch}{1.35}
    \begin{tabular}{@{}lcccc@{\hspace{2em}}cccc@{\hspace{2em}}cccc@{}}
        \toprule
         & \multicolumn{4}{c}{Soderblom $Z$ prior fit} & \multicolumn{4}{c}{Flat $Z$ prior fit} & \multicolumn{4}{c}{Flat $-$ Soderblom} \\
        \cmidrule(lr){2-5}\cmidrule(lr){6-9}\cmidrule(lr){10-13}
        Star & Age & $10^3 Z$ & $M$ & $v_{\rm eq}$ & Age & $10^3 Z$ & $M$ & $v_{\rm eq}$ & $\Delta\tau$ & $10^3\Delta Z$ & $\Delta M$ & $\Delta v_{\rm eq}$ \\
         & (Myr) & & ($\mathrm{M}_\odot$) & (km\,s$^{-1}$) & (Myr) & & ($\mathrm{M}_\odot$) & (km\,s$^{-1}$) & (Myr) & & ($\mathrm{M}_\odot$) & (km\,s$^{-1}$) \\
        \midrule
        V624\,Tau & $116.9^{+7.7}_{-8.2}$ & $17.87^{+0.17}_{-0.17}$ & $1.696^{+0.005}_{-0.005}$ & $48.1^{+0.2}_{-0.2}$ & $110.9^{+9.9}_{-10.4}$ & $17.92^{+0.22}_{-0.21}$ & $1.690^{+0.006}_{-0.006}$ & $48.1^{+0.2}_{-0.2}$ & $-6.0$ & $+0.04$ & $-0.007$ & $\phantom{+}0.0$ \\
        V647\,Tau & $113.7^{+9.3}_{-9.6}$ & $19.23^{+0.22}_{-0.22}$ & $1.721^{+0.006}_{-0.006}$ & $40.3^{+0.2}_{-0.2}$ & $110.2^{+12.2}_{-13.2}$ & $19.41^{+0.29}_{-0.30}$ & $1.719^{+0.007}_{-0.007}$ & $40.4^{+0.3}_{-0.3}$ & $-3.5$ & $+0.18$ & $-0.003$ & $+0.1$ \\
        V1187\,Tau & $112.6^{+17.5}_{-17.4}$ & $17.73^{+0.31}_{-0.35}$ & $1.665^{+0.011}_{-0.010}$ & $40.4^{+0.3}_{-0.3}$ & $107.7^{+21.9}_{-21.7}$ & $17.76^{+0.39}_{-0.43}$ & $1.663^{+0.013}_{-0.013}$ & $40.7^{+0.3}_{-0.3}$ & $-4.9$ & $+0.03$ & $-0.002$ & $+0.3$ \\
        HD\,23157 & $114.6^{+13.8}_{-12.2}$ & $18.88^{+0.14}_{-0.16}$ & $1.514^{+0.007}_{-0.006}$ & $76.7^{+0.2}_{-0.2}$ & $112.3^{+14.9}_{-13.6}$ & $18.86^{+0.14}_{-0.16}$ & $1.514^{+0.008}_{-0.006}$ & $76.6^{+0.3}_{-0.2}$ & $-2.4$ & $-0.02$ & $-0.001$ & $-0.2$ \\
        HD\,23325 & $107.0^{+9.3}_{-10.6}$ & $18.06^{+0.18}_{-0.17}$ & $1.639^{+0.008}_{-0.009}$ & $89.1^{+0.4}_{-0.5}$ & $99.5^{+12.0}_{-13.4}$ & $18.20^{+0.21}_{-0.21}$ & $1.629^{+0.010}_{-0.010}$ & $89.3^{+0.5}_{-0.5}$ & $-7.6$ & $+0.14$ & $-0.010$ & $+0.2$ \\
        \midrule
        Cluster & $113.3^{+4.5}_{-4.5}$ & --- & --- & --- & $108.5^{+5.5}_{-5.7}$ & --- & --- & --- & $-4.9$ & --- & --- & --- \\
        \bottomrule
    \end{tabular}
\end{table*}

\bsp
\end{document}